\documentclass[fleqn,usenatbib]{mnras}

\usepackage{newtxtext,newtxmath}
\usepackage{booktabs}
\usepackage{multirow}

\usepackage[T1]{fontenc}

\DeclareRobustCommand{\VAN}[3]{#2}
\let\VANthebibliography\thebibliography
\def\thebibliography{\DeclareRobustCommand{\VAN}[3]{##3}\VANthebibliography}

\usepackage{graphicx}	
\usepackage{amsmath}	

\newcommand{\apspr}{Astrophys. Space Phys. Res.}

\title[Generalized solution for accretion flows]{A generalized solution for advection-dominated accretion flow, standard disc, and slim disc}

\author[M. Liu, B. Liu, Y. Wang, H. Cheng, and W. Yuan]{
Mingjun Liu,$^{1,2}$
B. F. Liu,$^{1,2}$\thanks{E-mail: bfliu@nao.cas.cn}
Yilong Wang,$^{1,2}$
Huaqing Cheng$^{1}$
and Weimin Yuan$^{1,2}$
\\
$^{1}$National Astronomical Observatories, Chinese Academy of Sciences, 20A Datun Road, Beijing 100101, People's Republic of China\\
$^{2}$School of Astronomy and Space Science, University of Chinese Academy of Sciences, 19A Yuquan Road, Beijing 100049, People's Republic of China
}

\date{Accepted XXX. Received YYY; in original form ZZZ}

\pubyear{\the\year{}}

\begin{document}
\label{firstpage}
\pagerange{\pageref{firstpage}--\pageref{lastpage}}
\maketitle

\begin{abstract}
Aiming at a general description of four basic solutions describing the accretion processes, i.e., the Shakura-Sunyaev thin disc (SSD), the Shapiro-Lightman-Eardley solution (SLE), the slim disc, and the advection-dominated accretion flow (ADAF), we present generalized axisymmetric height-averaged equations, where the entropy advection, the radiation pressure, and photon trapping effect are all included self-consistently. Our generalized solution can reproduce the ADAF, SLE, SSD, and slim disc branches in a wide range of accretion rates from sub- to super-Eddington accretion. An S-curve in the $\dot{m}-\Sigma$ plane is also reproduced, representing the SSD branch, the radiation-pressure-dominant branch, and the slim disc branch. The solution gives a natural transition between SSD and slim disc when photon trapping occurs in the accretion flow, producing a radially hybrid SSD-slim disc structure in a certain range of accretion rates. The coexistence of ADAF and SSD below a critical accretion rate is clearly shown with distinct advection fraction of accretion energy. We also present the luminosity, radiation efficiency, and spectrum from the generalized solutions for a large range of accretion rates in stellar mass black holes.
\end{abstract}

\begin{keywords}
accretion, accretion discs -- hydrodynamics -- stars: black holes -- X-rays: general
\end{keywords}



\section{Introduction} \label{sec:intro}

The study of accretion dates back to the early investigations on spherically symmetric accretion, i.e., the Bondi accretion \citep[e.g.,][]{1940MNRAS.100..500A, 1944MNRAS.104..273B, 1952MNRAS.112..195B}. Since then, more realistic models have been proposed and four basic solutions of the accretion flow have been discovered, namely the standard disc \citep[SSD, e.g.,][]{1973A&A....24..337S,1973blho.conf..343N,1974MNRAS.168..603L}, the Shapiro-Lightman-Eardley solution \citep[SLE,][]{1976ApJ...204..187S}, the slim disc \citep[e.g.,][]{1977ApJ...215..265K,1978MNRAS.184...53B,1979MNRAS.187..237B,1982ApJ...253..873B,1988ApJ...332..646A,1993ApJ...412..254C,2005ApJ...628..368O}, and the advection-dominated accretion flow \citep[ADAF, e.g.,][]{1977ApJ...214..840I,1982Natur.295...17R,1994ApJ...428L..13N,1995ApJ...444..231N, 1995ApJ...452..710N,1995ApJ...438L..37A,1995ApJ...443L..61C}.

While the four solutions achieve great success in interpreting observations in various systems, they are obtained under specific conditions. The SSD solution describes the radiatively efficient, cold accretion flow at sub-Eddington accretion rate interpreting phenomena such as the high/soft state of X-ray binaries \citep[XRBs, e.g.][and references therein]{2006ARA&A..44...49R} and the big blue bump in active galactic nuclei \citep[AGNs, e.g.,][]{1982ApJ...254...22M,1983ApJ...268..582M,1987ApJ...321..305C,1994ApJS...95....1E}. The SLE solution also exists in low accretion rate but is thermally unstable \citep{1978ApJ...221..652P}, thereby it is not considered in real astrophysical systems. The ADAF solution describes the optically thin, hot accretion flow at low accretion rates and is believed to exist in low-luminosity AGNs, and XRBs in the hard and quiescent states \citep[see the review of][and references therein]{2014ARA&A..52..529Y}. The slim disc is expected to describe the super-Eddington accretion where the photons are trapped in matters, in systems such as the narrow-line Seyfert 1 galaxies \citep[e.g.,][]{1996A&A...305...53B,2000PASJ...52..499M,2012MNRAS.420.1825J,2012MNRAS.422.3268J,2012MNRAS.425..907J}, tidal disruption events \citep[TDEs, e.g.,][]{1988Natur.333..523R,1999A&A...343..775K,2018ApJ...859L..20D,2020ApJ...891..121L}, and ultraluminous X-ray sources \citep[ULXs, e.g.,][]{2011NewAR..55..166F,2017ARA&A..55..303K}.

Accretion flow can evolve among different accretion modes. The AGNs and XRBs are observed in a wide range of Eddington ratios. The accretion rates in TDEs can decline from super-Eddington regime to sub-Eddington regime on yearly scale. Any basic solution derived under specific conditions as mentioned above cannot fully describe the evolution of accreting system, especially for the transients powered by accretion. A general description of accretion flows to unify the solutions is essential.

For this purpose, one has to generalize the descriptions for the radiative transfer and the radiation pressure appropriate for both optically thin and thick flows, and the energy advection including entropy and trapping photons. In addition, the ion and electron temperatures for Coulomb-decoupled and coupled cases, the rotation deviating from Keplerian rotation, and possible radiation mechanisms are expected to be included in a generalized model. Attempts to unify the solutions have been made since the 1990s, as summarized in Table \ref{tab:model}. Specifically, at the early 1990s, \citet{1991ApJ...380...84W,1991ApJ...376..746L,1994MNRAS.266..386L} unified the SLE and SSD solutions with the phenomenological bridge formula for radiative transfer presented by \citet{1991ApJ...380...84W}. Then, taking a more physical description of the radiative transfer generalized for both optically thin and thick accretion flows \citep{1990ApJ...351..632H}, \citet{1995ApJ...452..710N} successfully unified the ADAF, the SLE, and the gas-pressure-supported SSD with their self-similar solution \citep{1994ApJ...428L..13N}. To include the slim disc, \citet{1995ApJ...438L..37A} treated the accretion flows with either radiation pressure or gas pressure. Then \citet{1995ApJ...443L..61C} made more improvements by introducing a generalized description of radiation pressure to unify these four solutions on the basis of Keplerian rotation and one-temperature flows. After that, a substantial number of works made great efforts on specific situations via construct models utilizing a part of generalized treatments \citep[e.g.][]{1996ApJ...471..762A,1996ApJ...465..312E,1996PASJ...48...77H,1997ApJ...482..400E,2000ApJ...538..295M,2001ApJ...549..100P,2001MNRAS.324..119Y,2003ApJ...594L..99Y,2009ApJ...697...16O,2011ApJ...733..112J,2023ApJ...954..150H}. There still lacks a highly self-consistent unified description of accretion flow.

In this work, we properly treat the radiation transfer, the total pressure, the energy advection, etc., as is in the last line of Table \ref{tab:model}. We present an algebraic approach for the generalized equations describing the accretion flows, making use of  the self-similar approximation. The solutions reproduce the ADAF, SLE, SSD, and slim disc at the corresponding parameter space and are expected to be more accurate at the crossing parameter space between the four solutions. The model is illustrated in Section ~\ref{sec:model}. The numerical results are presented in Section ~\ref{sec:results} and are discussed in Section ~\ref{sec:discussion}. The main conclusions are summarized in Section ~\ref{sec:conclusion}.

\begin{table*}
	\caption{A brief summary of the specific assumptions included in basic solutions and the unified descriptions of accretion flow}
	\label{tab:model}
	\resizebox{\textwidth}{!}{
	\begin{tabular}{lccccccc}
	    \toprule
		\multirow{2}{*}[-1ex]{Models} & \multirow{2}{*}[-1ex]{Rotation} & \multicolumn{2}{c}{Advection} & \multicolumn{2}{c}{Radiation} & \multirow{2}{*}[-1ex]{Pressure$^\dagger$} & \multirow{2}{*}[-1ex]{Temperature} \\\cmidrule(r){3-4}\cmidrule(r){5-6}
		\rule[-1ex]{0pt}{3.5ex} & & Entropy & Trapping & Transfer & Mechanism & & \\ 
		\midrule
		Bondi accretion  & \multirow{2}{*}[-1ex]{/} & \multirow{2}{*}[-1ex]{$\checkmark$} & \multirow{2}{*}[-1ex]{ $\times$} & \multirow{2}{*}[-1ex]{/} & \multirow{2}{*}[-1ex]{/} & \multirow{2}{*}[-1ex]{$p_\mathrm{g}$} & \multirow{2}{*}[-1ex]{$T_\mathrm{i}=T_\mathrm{e}$} \\
		{\citep[e.g.,][]{1952MNRAS.112..195B}}\\
	    SSD & \multirow{2}{*}[-1ex]{$\Omega=\Omega_\mathrm{K}$} & \multirow{2}{*}[-1ex]{$\times$} & \multirow{2}{*}[-1ex]{ $\times$} & \multirow{2}{*}[-1ex]{$\tau_\mathrm{eff}>1$} & \multirow{2}{*}[-1ex]{/} & \multirow{2}{*}[-1ex]{$p_\mathrm{g}+p_\mathrm{r,eq}$} & \multirow{2}{*}[-1ex]{$T_\mathrm{i}=T_\mathrm{e}$} \\
	    {\citep[e.g.,][]{1973A&A....24..337S}} \\
		SLE & \multirow{2}{*}[-1ex]{$\Omega=\Omega_\mathrm{K}$} & \multirow{2}{*}[-1ex]{$\times$} & \multirow{2}{*}[-1ex]{ $\times$} & \multirow{2}{*}[-1ex]{$\tau_\mathrm{eff}<1$} & \multirow{2}{*}[-1ex]{$q_\mathrm{br}+q_\mathrm{C}$} & \multirow{2}{*}[-1ex]{$p_\mathrm{g}$} & \multirow{2}{*}[-1ex]{$T_\mathrm{i}\gtrsim T_\mathrm{e}$} \\
		{\citep[e.g.,][]{1976ApJ...204..187S}} \\
		ADAF & \multirow{2}{*}[-1ex]{$\Omega\lesssim\Omega_\mathrm{K}$} & \multirow{2}{*}[-1ex]{$\checkmark$} & \multirow{2}{*}[-1ex]{ $\times$} & \multirow{2}{*}[-1ex]{$\tau_\mathrm{eff}<1$} & \multirow{2}{*}[-1ex]{$q_\mathrm{br}+q_\mathrm{syn}+q_\mathrm{C}$} & \multirow{2}{*}[-1ex]{$p_\mathrm{g}+p_\mathrm{m}$} & \multirow{2}{*}[-1ex]{$T_\mathrm{i}\gtrsim T_\mathrm{e}$} \\
		{\citep[e.g.,][]{1994ApJ...428L..13N,1995ApJ...452..710N}} \\
		Slim disc & \multirow{2}{*}[-1ex]{$\Omega\lesssim\Omega_\mathrm{K}$} & \multirow{2}{*}[-1ex]{$\checkmark$} & \multirow{2}{*}[-1ex]{ $\checkmark$} & \multirow{2}{*}[-1ex]{$\tau_\mathrm{eff}>1$} & \multirow{2}{*}[-1ex]{/} & \multirow{2}{*}[-1ex]{$p_\mathrm{g}+p_\mathrm{r,eq}$} & \multirow{2}{*}[-1ex]{$T_\mathrm{i}=T_\mathrm{e}$} \\
		{\citep[e.g.,][]{1988ApJ...332..646A}} \\
		\midrule
		SSD+SLE & \multirow{2}{*}[-1ex]{$\Omega=\Omega_\mathrm{K}$} & \multirow{2}{*}[-1ex]{$\times$} & \multirow{2}{*}[-1ex]{ $\times$} & \multirow{2}{*}[-1ex]{$\tau_\mathrm{eff}>1$ or $<1$} & \multirow{2}{*}[-1ex]{$q_\mathrm{br}+q_\mathrm{C}$} & \multirow{2}{*}[-1ex]{$p_\mathrm{g}+p_\mathrm{r}$} & \multirow{2}{*}[-1ex]{$T_\mathrm{i}\gtrsim T_\mathrm{e}$} \\
		{\citep[e.g.,][]{1991ApJ...380...84W}} \\
		ADAF+SLE+SSD & \multirow{2}{*}[-1ex]{$\Omega\lesssim\Omega_\mathrm{K}$} & \multirow{2}{*}[-1ex]{$\checkmark$} & \multirow{2}{*}[-1ex]{ $\times$} & \multirow{2}{*}[-1ex]{Any $\tau_\mathrm{eff}$} & \multirow{2}{*}[-1ex]{$q_\mathrm{br}+q_\mathrm{syn}+q_\mathrm{C}$} & \multirow{2}{*}[-1ex]{$p_\mathrm{g}+p_\mathrm{m}+p_\mathrm{r}$} & \multirow{2}{*}[-1ex]{$T_\mathrm{i}=T_\mathrm{e}$} \\
		{\citep[][]{1996ApJ...465..312E}} \\
		ADAF+SSD & \multirow{2}{*}[-1ex]{ $\Omega\lesssim\Omega_\mathrm{K}$} & \multirow{2}{*}[-1ex]{$\checkmark$} & \multirow{2}{*}[-1ex]{$\checkmark$} & \multirow{2}{*}[-1ex]{$\tau_\mathrm{eff}>1$ or $<1$} & \multirow{2}{*}[-1ex]{$q_\mathrm{br}$} & \multirow{2}{*}[-1ex]{\ $p_\mathrm{g}+p_\mathrm{r}$} & \multirow{2}{*}[-1ex]{ $T_\mathrm{i}=T_\mathrm{e}$} \\
		{\citep[e.g.,][]{1996PASJ...48...77H}} \\
		Unified description & \multirow{2}{*}[-1ex]{$\Omega=\Omega_\mathrm{K}$} & \multirow{2}{*}[-1ex]{$\checkmark$} & \multirow{2}{*}[-1ex]{ $\checkmark$} & \multirow{2}{*}[-1ex]{$\tau_\mathrm{eff}>1$ or $<1$} & \multirow{2}{*}[-1ex]{$q_\mathrm{br}+q_\mathrm{C}$} & \multirow{2}{*}[-1ex]{$p_\mathrm{g}+p_\mathrm{r}$} & \multirow{2}{*}[-1ex]{$T_\mathrm{i}=T_\mathrm{e}$} \\
		{\citep[][]{1995ApJ...443L..61C}} \\
		\multirow{2}{*}[0ex]{This work} & \multirow{2}{*}[0ex]{$\Omega\lesssim\Omega_\mathrm{K}$} & \multirow{2}{*}[0ex]{$\checkmark$} & \multirow{2}{*}[0ex]{$\checkmark$} & \multirow{2}{*}[0ex]{Any $\tau_\mathrm{eff}$} & \multirow{2}{*}[0ex]{$q_\mathrm{br}+q_\mathrm{syn}+q_\mathrm{C}$} & \multirow{2}{*}[0ex]{$p_\mathrm{g}+p_\mathrm{m}+p_\mathrm{r}$} & \multirow{2}{*}[0ex]{$T_\mathrm{i}\gtrsim  T_\mathrm{e}$} \\
		\\
		\bottomrule
	\end{tabular}}
    \footnotesize{$\dagger$ $p_{\rm r}$ stands for the generalized form of radiation pressure, i.e., the last formula of Eq.(\ref{eq:p}), and $p_{\rm r, eq}$ stands for $aT^4/3$.}
\end{table*}
 
\section{Model} \label{sec:model}

\subsection{Generalized equations describing the accretion flows} \label{subsec:equation}

The structure of a steady, axisymmetric, non-relativistic accretion flow is described by the continuity equation, radial momentum equation, azimuthal momentum equation, vertical hydrostatic equilibrium equation, and energy equations, supplemented by the equation of state (EOS). By solving these equations, the density $\rho$, the pressure $p$, the temperature $T_i$ and $T_e$, the radial velocity $\upsilon$, the angular velocity $\Omega$, and the vertical scale height $H$ can be obtained for a given mass of the accreting object $M$, the accretion rate $\dot{M}$, the distance $R$, the viscous parameter $\alpha$, and magnetic field if exist. We summarize the equations as follows. 
\begin{align}
    &\dot{M}=-4\pi \rho RH\upsilon,\label{eq:continuity}\\
    &\upsilon\frac{\text d\upsilon}{\text dR}-\Omega^2R=-\Omega^2_{\text K}R-\frac{1}{\rho}\frac{\text dp}{\text dR},\label{eq:radial}\\
    &\rho RH\upsilon\frac{\text d \left(\Omega R^2\right)}{\text dR}=\frac{\text d}{\text dR}\left(\nu\rho R^3H\frac{\text d\Omega}{\text dR}\right),\label{eq:angular}\\
    &\frac{p}{\rho H}=\frac{2}{5}\Omega_\mathrm{K}^2H,\label{eq:vertical}\\
    &\rho\upsilon\frac{\text de_\mathrm{int}}{\text dR}-\upsilon\frac{p}{\rho}\frac{\text d\rho}{\text dR}=\nu\rho \left(R\frac{\text d\Omega}{\text dR}\right)^2-\frac{F_{\rm rad}}{H},\label{eq:energy}
\end{align}
where $G$ is the gravitational constant, $\Omega_\mathrm{K}\equiv (GM/R^3)^{1/2}$ is the Keplerian angular velocity, $\nu=\alpha c_\mathrm{s}^2/\Omega_\mathrm{K}$ is the kinematic viscosity in connection with viscosity parameter $\alpha$ and the effective isothermal sound speed $c_\mathrm{s}\equiv \left(p/\rho\right)^{1/2}$, $e_{\rm int}$ is the internal energy per unit mass given in Eq.(\ref{eq:int}), $F_{\rm rad}$ is the radiation flux which is a function of temperature and optical depth. To describe the scale height of geometrically thick flow a factor of 2/5 is adopted in Eq.(\ref{eq:vertical}) \citep{1995ApJ...444..231N,1997ApJ...476...49N}, which does not affect the geometrically thin solution as $H/R$ is anyhow very small.

The equation of state is generalized by including the gas pressure, the magnetic pressure, and radiation pressure,
\begin{equation}\label{eq:p}
    \begin{split}
        &p=p_{\rm g}+p_{\rm m}+p_{\rm r},\\
        &p_{\rm g}=p_{\rm i}+p_{\rm e}=n_\mathrm{i}k_\mathrm{B}T_\mathrm{i}+n_\mathrm{e}k_\mathrm{B}T_\mathrm{e},\\
        &p_{\rm m}=\frac{B^2}{24\pi}\equiv\frac{1-\beta}{\beta}p_{\rm g},\\
        &p_{\rm r}=\frac{F_\mathrm{rad}}{2c}\left(\tau+\frac{2}{\sqrt{3}}\right),
    \end{split}
\end{equation}
where $k_\mathrm{B}$ in the gas pressure $p_{\rm g}$ is the Boltzmann constant,  $n_\mathrm{i}=\rho/\mu_\mathrm{i}m_\mathrm{u}$ and $n_\mathrm{e}=\rho/\mu_\mathrm{e}m_\mathrm{u}$ are respectively the number density of ions and electrons with $m_{\rm u}$ the atomic mass unit, $\mu_\mathrm{i}=1.23$ and $\mu_\mathrm{e}=1.14$ for an assumed mass fraction of hydrogen and helium of $0.75$ and $0.25$, respectively. The magnetic field $B$ is parameterized by $\beta$ assuming magnetic pressure in partition with the gas pressure, $\beta\equiv p_\mathrm{g}/(p_\mathrm{g}+p_\mathrm{m})$. The radiation pressure $p_{\rm r}$ is generalized to be applicable for both the optically thin and optically thick accretion flows \citep[e.g.,][]{1990ApJ...351..632H,1995ApJ...443L..61C,1996ApJ...471..762A}, where $c$ is the speed of light, $\tau=\tau_\mathrm{es}+\tau_\mathrm{abs}$ is the total optical depth from the middle plane to the surface of accretion flow, the scattering optical depth is given as $\tau_\mathrm{es}=n_\mathrm{e}\sigma_\mathrm{T}H$ with $\sigma_\mathrm{T}$ the Thomson scattering cross-section and the absorption optical depth $\tau_\mathrm{abs}$ expressed in Eq.(\ref{eq:abs}). 
 
Similar to the pressure, the energy density is also composed of the thermal energy, the magnetic energy, and the radiation energy, of which the latter two are 3 times of the magnetic pressure \citep[tangled magnetic field, e.g.,][]{1997ApJ...482..400E,1998tbha.conf..148N} and radiation pressure \citep[e.g.,][]{1983JQSRT..30..395B,  2003MNRAS.338.1013S} respectively. Thus, the internal energy per unit mass, $e_\mathrm{int}$, can be expressed as,
\begin{equation}\label{eq:int}
        e_\mathrm{int}=\frac{1}{\gamma-1}\frac{p_\mathrm{g}}{\rho }+\frac{3p_\mathrm{m}}{\rho}+\frac{3p_\mathrm{r}}{\rho}\equiv \frac{1}{\Gamma_3-1}\frac{p}{\rho},
\end{equation}
where $\gamma=5/3$ is the ratio of specific heats for ideal gas and $\Gamma_3$ is the general adiabatic exponent by inclusion of the radiation pressure, magnetic pressure, and gas pressure \citep[e.g.,][]{1967aits.book.....C,1983psen.book.....C,1984oup..book.....M}.

In order to describe the decoupling of ions and electrons, as in the two-temperature ADAF and the SLE, an additional equation describing the energy balance for electrons is necessary to complete the equations of accretion flows. We assume heating via Coulomb collisions with ions is balanced by radiative cooling,  
\begin{equation}
    q_{\rm ie}=q_{\rm rad},
        \label{eq:electron}
\end{equation}
where $q^{\mathrm{ie}}$ is the energy transfer rate between ions and electrons (see Appendix \ref{app:model}), and $q_{\rm rad}$ is the height-averaged net radiation rate by electrons, $q_{\rm rad}\equiv F_{\rm rad}/H$.

The effective surface flux $F_\mathrm{rad}$ is obtained from the vertical radiative transfer \citep[see][]{1990ApJ...351..632H},
\begin{equation}
    F_\mathrm{rad}=4\sigma T_\mathrm{e}^4\left(\frac{3}{2}\tau+\sqrt{3}+\frac{1}{\tau_\mathrm{abs}}\right)^{-1},
    \label{eq:q_rad}
\end{equation}
where $\sigma$ is the Stefan-Boltzmann constant. 

Different from the Rosseland approximation adopted for optically thick SSD, Eq.(\ref{eq:q_rad}) has been derived from the Eddington approximation with the two-stream approximation for the upper boundary \citep[][]{1990ApJ...351..632H}. Thus, the flux expressed by Eq.(\ref{eq:q_rad}) is valid for both optically thin and optically thick accretion flows. In this case, the absorption coefficient, $\alpha_\mathrm{abs}$, is no longer the Rosseland mean absorption coefficient, but can be expressed by the emissivity from the Kirchhoff's law as a thermal equilibrium system, i.e. $\alpha_\mathrm{abs}=(q_\mathrm{br}+q_\mathrm{syn}+q_\mathrm{br,C}+q_\mathrm{syn,C})/4\sigma T_\mathrm{eff}^4$, which gives \citep[see also ][]{1995ApJ...452..710N},
\begin{equation}
    \tau_\mathrm{abs}=\frac{\left(q_\mathrm{br}+q_\mathrm{syn}+q_\mathrm{br,C}+q_\mathrm{syn,C}\right)H}{4\sigma T_\mathrm{e}^4},
    \label{eq:abs}
\end{equation}
where $q_\mathrm{br}$, $q_\mathrm{syn}$, $q_\mathrm{br,C}$, and $q_\mathrm{syn,C}$ are the emissivity of bremsstrahlung, synchrotron, bremsstrahlung self-Compton, and synchrotron self-Compton processes, respectively. The detailed description of each emissivity is shown in Appendix \ref{app:model}.

Combined with Eq.(\ref{eq:abs}), Eq(\ref{eq:q_rad}) appropriately describes the emergent radiation flux. In the extreme cases, Eq(\ref{eq:q_rad}) gives $F_{\rm rad}\approx (q_\mathrm{br}+q_\mathrm{syn}+q_\mathrm{br,C}+q_\mathrm{syn,C})H$ for $\tau \ll 1$ and $F_\mathrm{rad}\approx 8\sigma T_\mathrm{e}^4/3\tau$ for $\tau \gg 1$, consistent with that in the ADAF and the SSD respectively. Note that the net radiation rate $q_{\rm rad}=F_{\rm rad}/H$ does not equal to local total emissivity $q_\mathrm{br}+q_\mathrm{syn}+q_\mathrm{br,C}+q_\mathrm{syn,C}$ in the optically thick case.

Eqs.(\ref{eq:continuity})--(\ref{eq:p}) and (\ref{eq:electron}) are the generalized equations describing the accretion flows, from which all possible solutions can, in principle, be derived including the basic four solutions under specific conditions.

\subsection{Algebraic approach by virtue of self-similar solution}

Solving the generalized, non-linear equations is complicated, in particular, when various radiation and absorption processes are taken into account. On the other hand, the self-similar solution provides the simple and convenient ways to reveal the physical properties of the solutions and can be verified to be a good approximation (see Appendix \ref{app:solution}). For this purpose, we adopt the self-similar forms of the $\Omega$, $\upsilon$, and $c_\mathrm{s}^2$, and introduce the advection fraction $f$ and generalized adiabatic exponent $\Gamma_3$, similar to \citet{1994ApJ...428L..13N, 1995ApJ...452..710N}. In this way, the radiation flux $F_{\rm rad}$ doesn't explicitly occur to the energy  Eq.(\ref{eq:energy}), which can be rewritten as
\begin{equation}\label{eq:energy-re}
    \rho\upsilon\frac{\text d}{\text dR}\left(\frac{1}{\Gamma_3-1}\frac{p}{\rho}\right)-f\upsilon\frac{p}{\rho}\frac{\text d\rho}{\text dR}=f\nu\rho \left(R\frac{\text d\Omega}{\text dR}\right)^2.
\end{equation}

Therefore, the self-similar velocities have the form as
\begin{equation}\label{eq:omega}
    \begin{split}
        &\upsilon=-\frac{\left(5+2\epsilon^\prime\right)}{3\alpha}g\left(\alpha,\epsilon^\prime\right)v_\mathrm{K}\equiv-\alpha c_1\upsilon_\mathrm{K},\\
        &\Omega=\left[\frac{2\epsilon^\prime\left(5+2\epsilon^\prime\right)}{9\alpha^2}g\left(\alpha,\epsilon^\prime\right)\right]^{1/2}\Omega_\mathrm{K}\equiv c_2\Omega_\mathrm{K},\\
        &c_\mathrm{s}^2=\frac{2\left(5+2\epsilon^\prime\right)}{9\alpha^2}g\left(\alpha,\epsilon^\prime\right)\upsilon_\mathrm{K}^2\equiv c_3\upsilon_\mathrm{K}^2,
    \end{split}
\end{equation}
with
\vspace{-\baselineskip}

\begin{equation*}
    \begin{aligned}
     &\upsilon_\mathrm{K}=(GM/R)^{1/2},\\
     &g\left(\alpha,\epsilon^\prime\right)\equiv\left[1+\frac{18\alpha^2}{\left(5+2\epsilon^\prime\right)^2}\right]^{1/2}-1,\\    
     &\epsilon^\prime\equiv\frac{1}{f}\left(\frac{5/3-\Gamma_3}{\Gamma_3-1}+1-f\right).
    \end{aligned}
\end{equation*}
Compared with that of \citet{1994ApJ...428L..13N}, the above expression differs only in $\epsilon^\prime$ due to two reasons. First, we re-define the advection factor, $f$, leading to the corresponding change in the expression of $\epsilon'$. In contrast to its definition as the ratio of advected entropy and viscous heating in \citet{1994ApJ...428L..13N},  $f$ in this work is defined as the true advected energy fraction, that is, 
\begin{equation}
 f\equiv\frac{q_{\rm int}}{q_{\rm c}+q_{\rm vis}}, 
\end{equation}
where $q_{\rm int}\equiv {\rho\upsilon\text de_\mathrm{int}/\text dR}$ is the advected energy rate, $q_{\rm c}\equiv\upsilon\left(p/\rho\right)\text d\rho/\text dR$ is the pressure work rate, and $q_{\rm vis}\equiv \nu\rho \left(R\text d\Omega/\text dR\right)^2 $ is viscous heating rate, which are all shown in the energy equation (\ref{eq:energy}). Note that the denominator ($q_{\rm c}+q_{\rm vis}$) equals to the changing rate of mechanic energy, i.e. the release rate of accretion energy, thus the new definition of $f$  represents the advected  fraction of accretion energy, which is dependent on the pressure, density, and temperature.

The second difference is that $\gamma$ for the ADAF \citep{1994ApJ...428L..13N} is replaced by the general adiabatic exponent $\Gamma_3$ since the radiation pressure is included in this work, in addition to the magnetic pressure and gas pressure for the ADAF.

By using the self-similar solutions $\upsilon$ and $c_s^2$, the density $\rho$ and scale height $H$ can be solved through the continuity equation (\ref{eq:continuity}) and the vertical hydrostatic equilibrium equation (\ref{eq:vertical}), provided that $f$ and $\Gamma_3$ are known. The solutions are listed below,
\begin{equation}\label{eq:prhoHq}
    \begin{split}
        &H=\left(\frac{5}{2}\right)^{1/2}\frac{c_\mathrm{s}}{\Omega_\mathrm{K}},\\
        &\rho=\frac{\dot{M}}{4\pi RH|\upsilon|},\\
        &\Sigma=2\rho H,\\
        &p=\rho c_\mathrm{s}^2,\\
        &q_{\rm vis}=\frac{3\epsilon^\prime\rho|\upsilon|c_\mathrm{s}^2}{2R}.
    \end{split}
\end{equation}

The next task is to solve the temperature $T_\mathrm{i}$, $T_\mathrm{e}$, $f$, and $\Gamma_3$ from the state equation (Eq.\ref{eq:p}), the definition equation of $\Gamma_3$ (Eq.\ref{eq:int}), and the two energy equations for the total accretion flows (Eq.\ref{eq:energy}) and the electrons alone (Eq.\ref{eq:electron}), which are elucidated in the following subsection.

\subsection{Numerical method} \label{subsec:method}
Given the mass of central black hole, the accretion rate, the distance, the viscosity parameter and the magnetic parameter, the self-similar solution for velocities and the parameters $H$, $p$, $\rho$ and $q_{\rm vis}$ can be calculated out from Eqs. (\ref{eq:omega}) and (\ref{eq:prhoHq}) with presumed values for $f$ and $\Gamma_3$. By using these self-similar solutions and the detailed description of energy balance in Appendix \ref{app:model}, the temperature $T_\mathrm{i}$, $T_\mathrm{e}$, $\Gamma_3$, and $f$ can be solved from the energy equation (Eq.\ref{eq:energy-re}), the EOS (Eq.\ref{eq:p}), the definition of $\Gamma_3$ (Eq.\ref{eq:int}), and the energy equation of electrons (Eq.\ref{eq:electron}). In the computation, we technically introduce a new variable $\chi\equiv p_{\rm r}/p$ to replace $\Gamma_3$, with which $\Gamma_3$ can be expressed as a function of $\chi, \gamma$, and $\beta$ by combining Eqs.(\ref{eq:p}) and (\ref{eq:int}). Then the equations for solving $T_\mathrm{i}$, $T_\mathrm{e}$, $\chi$, and $f$ are summarized and rewritten as,
\begin{equation}\label{eq:fTiTe}
\begin{split}
    &(1-f)(q_{\rm vis}+q_{\rm c})=q_{\rm rad},\\
    &q_{\rm ie}=q_{\rm rad},\\
    &(1-\chi)p=\frac{n_\mathrm{i}k_\mathrm{B}T_\mathrm{i}+n_\mathrm{e}k_\mathrm{B}T_\mathrm{e}}{\beta},\\
    &\chi=\frac{F_\mathrm{rad}}{2cp}\left(\tau+\frac{2}{\sqrt{3}}\right).
\end{split}
\end{equation}

In the practical computation, we arbitrarily give initial values of $f$, $\chi$, $T_\mathrm{i}$, and $T_\mathrm{e}$, with which we can determine all the parameters as listed Eq.(\ref{eq:prhoHq}) and then determine $T_{\rm i}$, $T_e$, $\chi$, and $f$ by Eq.(\ref{eq:fTiTe}). The derived values for $f$ and $\chi$ are then compared with the presumed values, iterative computations continue until the presumed values for $f$ and $\chi$ equal to the derived ones, which are then taken as the true solution. 

To improve the efficiency of searching solutions and avoid missing solutions, we develop an ergodic-bisection method based on \citet{1995ApJ...452..710N,1996ApJ...465..312E}. Assuming the values of $f$, $\chi$, and $T_\mathrm{e}$, we can obtain the value of $T_\mathrm{i}$ from the EOS  (\ref{eq:p}). Then, we can verify whether the value of $T_\mathrm{e}$ satisfies the electron energy balance (\ref{eq:electron}). If not, we change the value of $T_\mathrm{e}$ by the dichotomy method. Once the Eq.(\ref{eq:electron}) is satisfied, we verify the definition of $\chi$. Similarly, we change the value of $\chi$ by the dichotomy method until the fourth equation in Eq.(\ref{eq:fTiTe}) is satisfied. Finally, with the $\chi$, $T_\mathrm{e}$, and $T_\mathrm{i}$, we check the energy equation (\ref{eq:energy}) for the assumed $f$. To avoid missing any solution, we vary $f$ to go through enough points from 0 to 1. For each value of $\dot{M}$, at most three solutions can be found, consistent with the analysis in Appendix \ref{app:stable}.

In addition, during the search of $f$, for each $f$ we can obtain the $T_\mathrm{i}$ and the error of energy balance, $q_\mathrm{rad}/(1-f)(q_\mathrm{vis}+q_\mathrm{c}) - 1$. From the slope of $q_\mathrm{rad}/(1-f)(q_\mathrm{vis}+q_\mathrm{c})-T_\mathrm{i}$ curve, we can determine the locally thermal stability of each solution.

\section{Results} \label{sec:results}

In this work, we adopt typical parameters, $\alpha=0.1$, $\beta=0.5$ (equivalent to $p_\mathrm{m}=p_\mathrm{g}$), $M=10M_\odot$ and perform the calculations for a large range of accretion rates in order to obtain the solutions covering ADAF, SLE, SSD, and slim disc regime. For convenience the mass of central black hole, the accretion rate, and radius are scaled in solar mass, Eddington rate, and Schwarzschild radius respectively, that is, $m=M/M_\odot$, $\dot m=\dot M/\dot M_{\rm Edd}$, $r=R/R_{\rm S}$ with $M_\odot=1.989\times 10^{33}\ {\rm g}$, $\dot{M}_\mathrm{Edd}=1.4\times 10^{18}M/M_\odot\, {\rm g/s}$, $R_\mathrm{S}=2.95\times 10^5 M/M_\odot\, {\rm cm}$.

\subsection{Regimes of the four solutions}\label{subsec:mdot}
\begin{figure*}
\centering 
    \includegraphics[width=\textwidth]{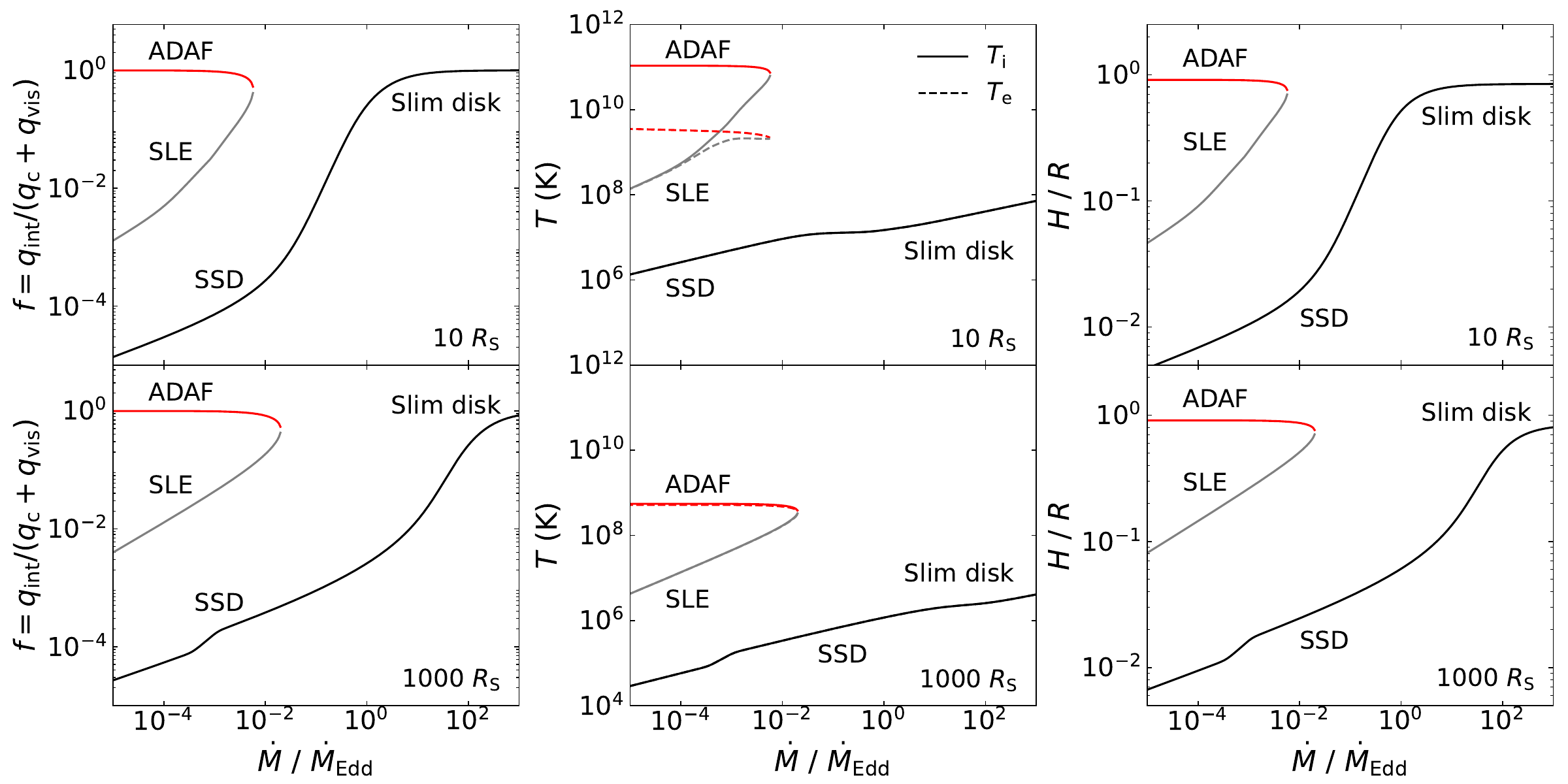}
    \caption {All possible solutions obtained from the generalized equations. The left, middle, and right panels display respectively the  advection fraction of energy, the ion and electron temperatures, and the scale height varying with the accretion rates from sub- to super-Eddington accretion rates for $m=10$, $\alpha=0.1$, and  $p_\mathrm{m}=p_\mathrm{g}$. The upper panels are for  
    $R=10R_\mathrm{S}$ and the lower panels are for $R=1000R_\mathrm{S}$.}
    \label{fig:properties}
\end{figure*}

The generalized solution reproduces the four patterns of accretion flows from sub- to super-Eddington accretion rates as shown in Figure \ref{fig:properties} and Figure \ref{fig:s_curve}. In Figure \ref{fig:properties}, the left panels display how the advection fraction of energy varies with the accretion rate. In the regime of low accretion rates, the ADAF, SLE, and SSD solutions clearly exist, of which the SLE approaches the ADAF at the upper critical accretion rate $\dot{m}_\mathrm{c}$ and converges to the SSD at very low accretion rates (out of the figure edge). The advection energy is dominant in ADAF, whereas it is small in SLE, leading to the unstable feature of SLE. In the regime of high accretion rates, the SSD changes from gas pressure supported to radiation pressure dominated disc and then transfers to a slim disc with the increase of the accretion rate, of which the slim disc extends up to super-Eddington rates and the SSD continues down to the regime of low accretion rate. These characteristics change somewhat with the distance to the black hole, as shown in the upper and lower panels. From $10R_\mathrm{S}$ to 1000 $R_\mathrm{S}$, the upper limit for the existence of ADAF and SLE increases and the transition from SSD to slim disc occurs at increasing accretion rates from $\sim1$ $\dot{M}_\mathrm{Edd}$ to $\sim10^2$ $\dot{M}_\mathrm{Edd}$. This is caused by less release of gravitational energy at larger distances.

The properties of the mid-plane temperature are displayed in the middle panels. At low accretion rates, there are three sets of solutions corresponding to the ADAF, SLE, and SSD. The ADAF solutions are always two-temperature due to the weak Coulomb coupling, the SLE solutions can also become two-temperature at the inner region when the accretion rate is close to $\dot{m}_\mathrm{c}$ and the SSD solutions are single-temperature. With the increase of the accretion rate, the temperatures in SSD rise first, then stay almost constant, and rise again, which correspond respectively to the gas-pressure-supported SSD, radiation-pressure-supported SSD, and slim disc stages. As the distance to the central black hole increases, the temperature of all solutions decreases and the ADAF solution gradually becomes single-temperature.

The scale height of accretion flow, as shown in the right panels of Figure \ref{fig:properties}, seems to well trace the advection fraction of energy. The ADAF solution is always geometrically thick as a consequence of energy advection. The SLE solution becomes geometrically thick near the $\dot{m}_\mathrm{c}$ due to its high temperature. The SSD is geometrically thin with $H/R<10^{-2}$ at low accretion rates, gradually puffs up with the increase of radiation pressure at high accretions, and transfers to geometrically thick slim disc as a result of overwhelming radiation pressure with photons trapped in the accretion flows. This effect is less obvious at larger distances where photon-trapping occurs at higher accretion rates.

\subsection{$\dot M-\Sigma$ relation of the generalized solutions} \label{subsec:s_curve}
\begin{figure*}
\centering 
    \includegraphics[width=\textwidth]{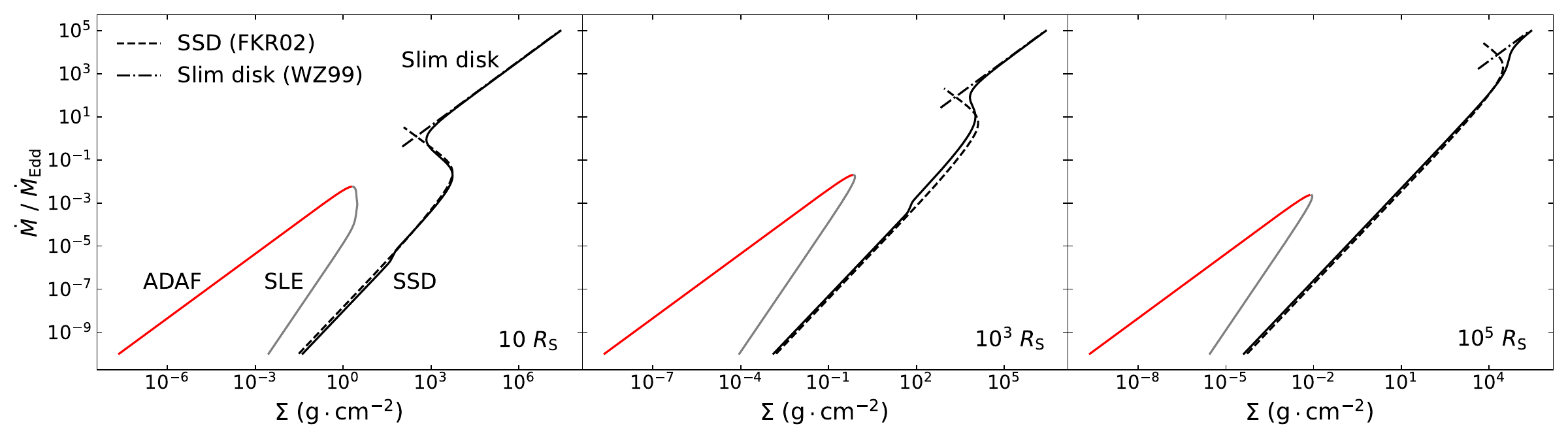}
    \caption{The thermal equilibrium curves of the generalized solutions composed of ADAF, SLE, SSD, and slim disc for $m = 10$, $\alpha=0.1$, and $p_\mathrm{m}=p_\mathrm{g}$ at 10 $R_\mathrm{S}$ (the left panel), $10^3$ $R_\mathrm{S}$ (the middle panel), and $10^5$ $R_\mathrm{S}$ (the right panel). For comparison, the thermal equilibrium curves from the SSD equations \citep[][FKR02]{2002apa..book.....F} and the self-similar solution of slim disc \citep[][WZ99]{1999ApJ...516..420W} are also plotted by black dashed line and black dot-dashed line, respectively. It should be noted that if the partial ionization is considered, a second S-curve at sub-Eddington accretion rate will develop.}
    \label{fig:s_curve}
\end{figure*}

The thermal equilibrium curves for the generalized solutions are shown in Figure \ref{fig:s_curve} for $10\ R_{\rm S}$, $10^3\ R_{\rm S}$, and $10^5\ R_{\rm S}$. In this $\dot{m}-\Sigma$ plane, the left curve is composed of the ADAF and the SLE, which completely overlap with that of \citet{1995ApJ...452..710N}. The right S-curve is composed of the gas-pressure-supported SSD, radiation-pressure-supported SSD, and slim disc. For verification of our generalized solution, we plot in Figure \ref{fig:s_curve} the SSD solution of equation (5.41) in \citet{2002apa..book.....F} with the inclusion of the magnetic field and the self-similar solution of slim disc of \citet{1999ApJ...516..420W}. Figure \ref{fig:s_curve} shows that the S-curve is well consistent with the solutions by \citet{2002apa..book.....F} and \citet{1999ApJ...516..420W} in a wide range of accretion rates. In particular, our generalized solution results in a smooth transition between the SSD and slim disc.

The ADAF, gas-pressure-supported SSD, and slim disc branches are thermal stable, while the SLE and radiation-pressure-supported SSD branches are thermally unstable, as analyzed in Appendix \ref{app:stable}. The formation of the S-curve can be understood as follows. Starting from the lower, gas-pressure-dominated SSD branch, the surface density increases with increasing accretion rate, and so does the temperature. Since the radiation pressure increases much more drastically than the gas pressure, approximately the fourth power of temperature in the optically thick case, the radiation pressure dominates over the gas pressure when the accretion rate increases to a certain value. From then on, the further increase of the accretion rate leads to the vertical expansion of the disc and the steep increase of radial velocity ($\upsilon\propto\dot M^2$), which results in a decrease of surface density as $\Sigma\propto \dot M/ \upsilon\propto\dot M^{-1}$. This forms the radiation pressure-dominant unstable branch. For the even higher accretion rate, the scattering depth is so high that photons produced deep in the accretion flow cannot escape before accreting to the black hole, i.e. photon trapping occurs. In this case, the radial gradient of pressure is comparable with centrifugal force, thus, the radial velocity  does not vary with the accretion rate ($\upsilon\propto\alpha c_\mathrm{s}\propto\dot{m}^0$). Therefore, the surface density rises with $\dot M$ again, that is the branch of stable, slim disc. At large distances, as shown in the middle panel of Figure \ref{fig:s_curve}, the release of accretion energy is not strong, hence the temperature and density are low. Thus, the radiation-pressure-dominant branch and slim disc branch only occur at very high accretion rates. At even larger distances, the radiation-pressure unstable branch shrinks and finally disappears, as shown in the right panels of Figure \ref{fig:s_curve} for $R\simeq 10^5 R_\mathrm{S}$. Photon-trapping  stabilizes the inner disc and eventually the whole disc when the accretion rate is higher than $10^4$ $\dot{M}_\mathrm{Edd}$.

\subsection{The radial structure of accretion flows}
\label{subsec:r}

\begin{figure*}
	\includegraphics[width=0.95\textwidth]{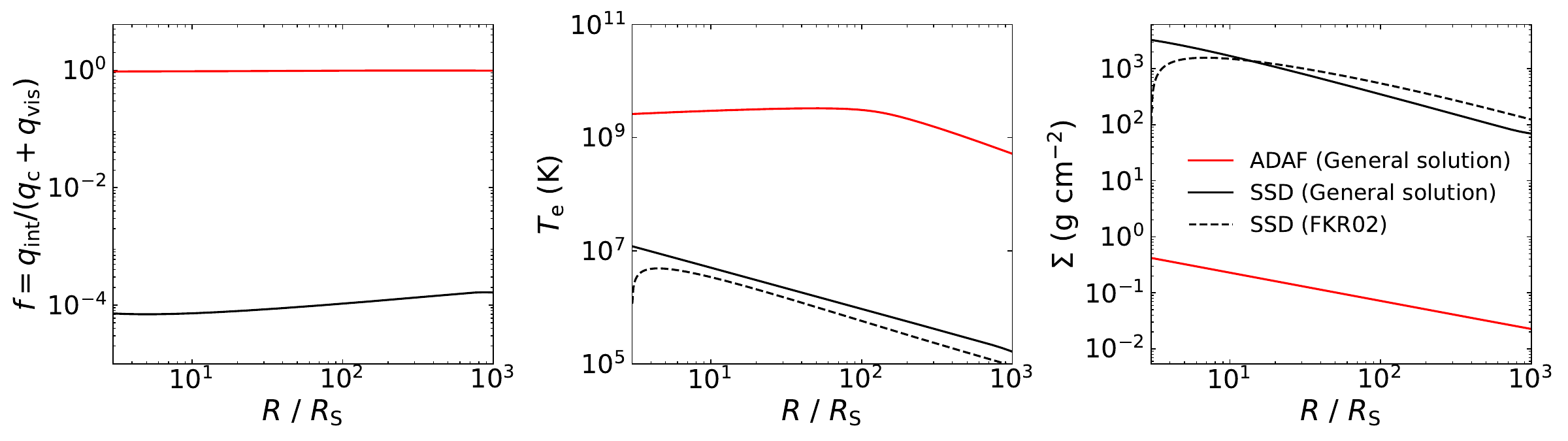}
    \caption{The radial profiles of the advection fraction of energy, electron temperatures, and surface density of the ADAF (red solid lines) and SSD (black solid lines) in generalized solution for $m=10$, $\alpha=0.1$, and $p_\mathrm{m}=p_\mathrm{g}$, at $0.001\ \dot{M}_\mathrm{Edd}$. For comparison, the SSD solutions of \citet[][FKR02]{2002apa..book.....F} are plotted in black dashed line.}
    \label{fig:radial_adaf}
\end{figure*}

\begin{figure}
	\includegraphics[width=\columnwidth]{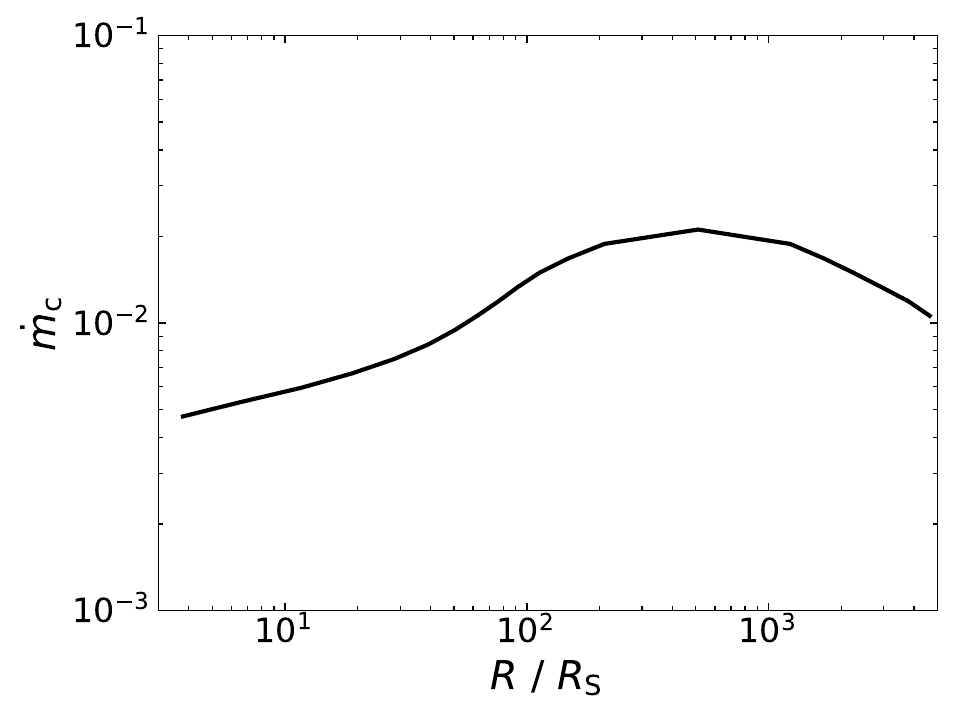}
    \caption{The critical accretion rate for existence of hot flows as a function of radius for $m=10$, $\alpha=0.1$, and $p_\mathrm{m}=p_\mathrm{g}$.}
    \label{fig:crit}
\end{figure}

\begin{figure*}
    \centering 
    \includegraphics[width=0.95\textwidth]{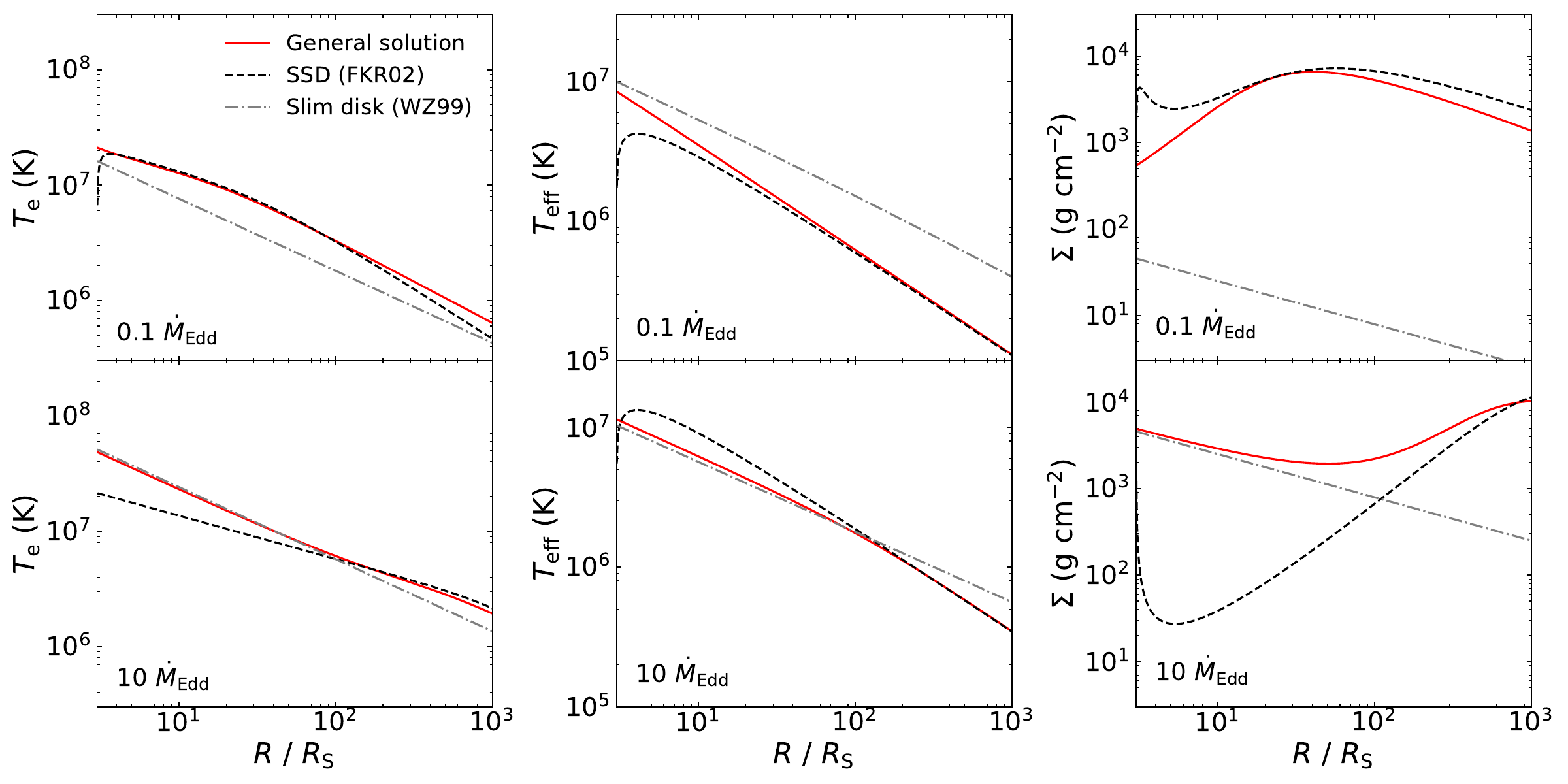}
    \caption{The radial profiles (red solid lines) of electron temperatures, effective temperature, and surface density of the generalized solution for $m=10$, $\alpha=0.1$, and $p_\mathrm{m}=p_\mathrm{g}$, at $0.1\ \dot{M}_\mathrm{Edd}$ (the upper panel) and $10\ \dot{M}_\mathrm{Edd}$ (the lower panel). For comparison, the SSD solutions of \citet[][FKR02]{2002apa..book.....F} and the slim disc solutions of \citet[][WZ99]{1999ApJ...516..420W} are also plotted in black dashed line and grey dot-dashed line, respectively.}
    \label{fig:radial}
\end{figure*}

The radial structure of our generalized solutions for the ADAF is essentially the same as the self-similar solution of \citet{1994ApJ...428L..13N,1995ApJ...452..710N} since the emission is so weak that the radiation pressure is negligible and no photon can be trapped. As shown in Figure \ref{fig:radial_adaf}, the advected energy faction of our ADAF solution almost remains constant. The electron temperature is nearly constant in the inner region and then decreases in outer region, which is consistent with the results of \citet{1995ApJ...452..710N}. The 
surface density has a power-law profile. Here we also display the distribution of the upper limit of the accretion rate for the existence of hot flows, $\dot{m}_\mathrm{c}$. As shown in Figure \ref{fig:crit}, $\dot{m}_\mathrm{c}$ increases with radius until $R \sim 10^3 R_\mathrm{S}$ and then decreases, consistent with  \citet{1995ApJ...452..710N,1996ApJ...465..312E,2023MNRAS.521.3237L}. The variation of $\dot m_c$ at $R \la 10^3 R_\mathrm{S}$ is a consequence of more efficient Coulomb collisions at smaller distances when a steady accretion flow is compressed into a smaller space, which leads to more efficient radiations through inverse Compton scatterings and hence the upper limit for a radiation-inefficient ADAF is reached at a relatively lower accretion rate. The decrease of $\dot{m}_\mathrm{c}$ at $R\ga 10^3 R_\mathrm{S}$ can be understood as the viscous heating is dominantly cooled by the bremsstrahlung radiation, which is more efficiently at larger distances. Detailed analysis is illustrated in Appendix \ref{app:stable}. With the inclusion of synchrotron and Compton cooling, our $\dot m_{\rm c}(R)$ is more precise than that of \citet{1995ApJ...438L..37A,1995ApJ...443L..61C,1996PASJ...48...77H}.

The radial structure of the generalized solutions for the SSD and slim disc are shown in Figure \ref{fig:radial_adaf} and Figure \ref{fig:radial}. The SSD and ADAF solutions distinctly exist below the $\dot{m}_\mathrm{c}$, of which the SSD smoothly extends to higher accretion rates. There is only the SSD solution for the whole radial region at an accretion rate of 0.1 Eddington rate, whereas the SSD transfers to the slim disc in the inner region as photon trapping occurs at an accretion rate of 10 Eddington rate, giving a hybrid radial structure. To verify the radial structure of our generalized solution, we numerically solve the equations in \citet{2002apa..book.....F} and \citet{1999ApJ...516..420W} as mentioned in Section \ref{subsec:s_curve} and plot the results in Figure \ref{fig:radial_adaf} and Figure \ref{fig:radial}. As can be seen from Figure \ref{fig:radial_adaf} and the upper panels of Figure \ref{fig:radial}, our generalized SSD solution is well consistent with the works of \citet{2002apa..book.....F}. The deviation of the effective temperature and surface density in the innermost region is caused by the free inner boundary in contrast to the zero shear stress in \citet{2002apa..book.....F}. The slight difference in the outermost region is due to different treatments on absorption involved in the radiation transfer. For super-Eddington accretion, photon trapping occurs in inner regions. As shown in the lower panels of Figure \ref{fig:radial}, our results are consistent with the slim disc solution by \citet{1999ApJ...516..420W} in the inner region and are consistent with the SSD solution by \citet{2002apa..book.....F} in the outer region. A smooth transition is shown between SSD and slim disc by the generalized solution.

Note that there is no intersection in the radial distribution of temperature and surface density between the SSD and ADAF (see Figure \ref{fig:radial_adaf}), which implies that the SSD extends down to the innermost stable circular orbit (ISCO) if the accretion in the outer region is via SSD. Therefore, a truncation of a thin disc to an ADAF in the inner region was ascribed to "strong ADAF" assumption, i.e., whenever an ADAF can be exist, the accretion pattern would be the ADAF. A more physical mechanism  is  the disc evaporation model \citep[e.g.,][]{, meyer1994, liu1999,2000A&A...361..175M}, in which the thin disc can be completely evaporated into an ADAF at low accretion rates. Both of the models predict the formation of the inner hot flows for the accretion rate below a critical value, which depends on the viscosity parameter roughly as $\dot m_{\rm c}\propto \alpha^2$. Such results  are confirmed by our generalized solutions (see also Eq. \ref{eq:mdotc}).

\subsection{The luminosity and radiation efficiency}

\begin{figure}
	\includegraphics[width=\columnwidth]{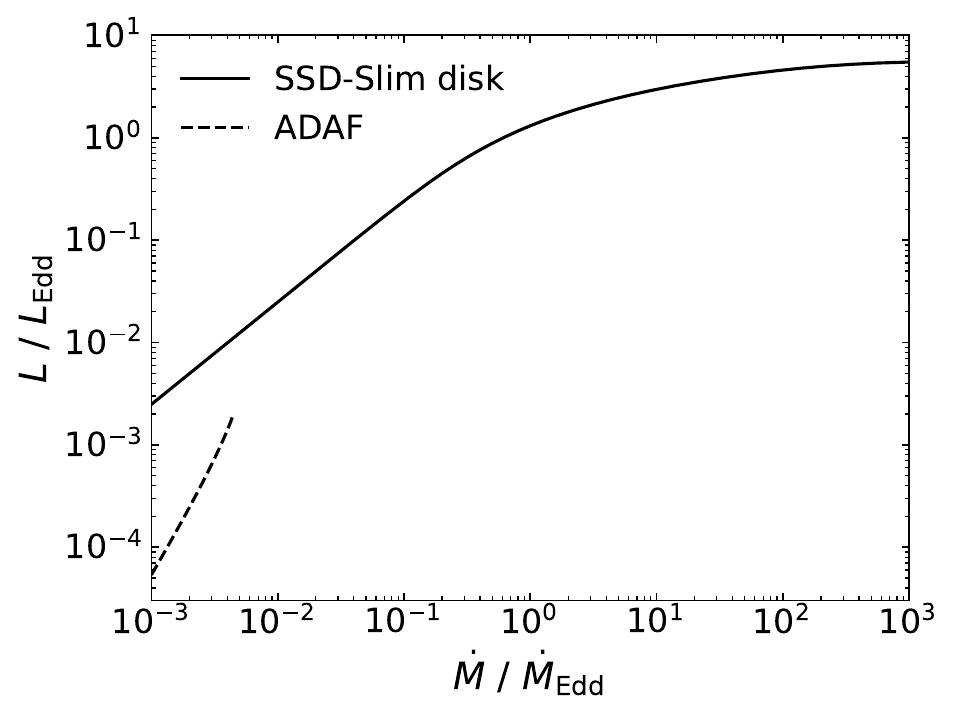}
    \caption{The variation of the bolometric luminosities as a function of the mass accretion rates of SSD-slim disc and the ADAF for $m=10$, $\alpha=0.1$, and $p_\mathrm{m}=p_\mathrm{g}$.}
    \label{fig:lum}
\end{figure}

\begin{figure}
	\includegraphics[width=\columnwidth]{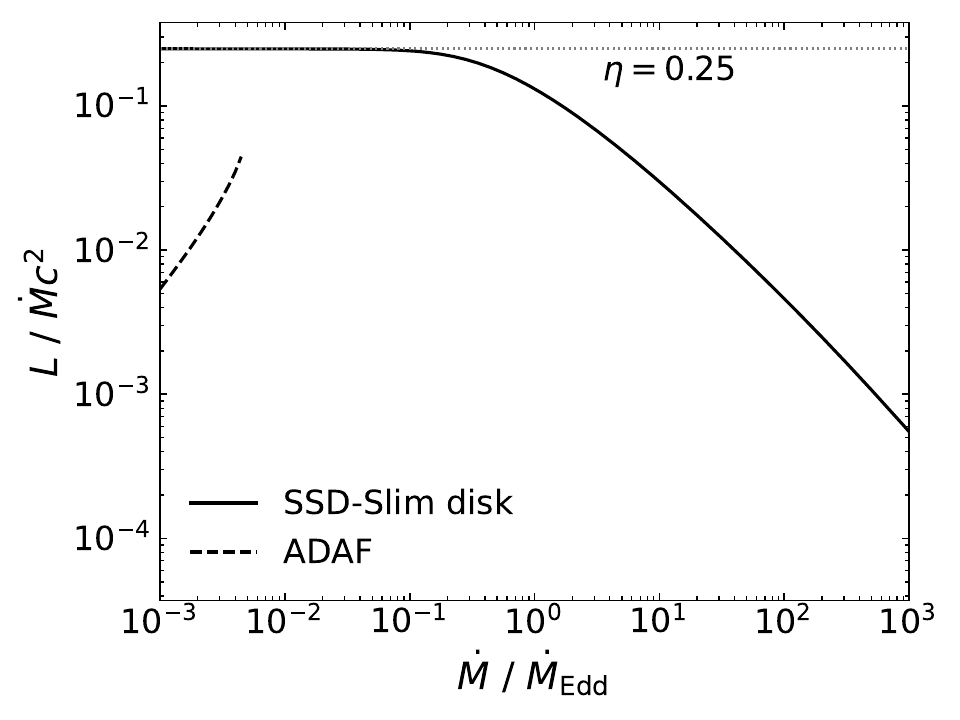}
    \caption{The variation of the radiation efficiency as a function of the mass accretion rates of SSD-slim disc and the ADAF for $m=10$, $\alpha=0.1$, and $p_\mathrm{m}=p_\mathrm{g}$. The radiation efficiency of SSD without zero torque condition at inner boundary is also marked by horizontal grey dotted line.}
    \label{fig:eff}
\end{figure}

The luminosities and radiation efficiencies of ADAF, SSD, and slim disc are shown in Figure \ref{fig:lum} and Figure \ref{fig:eff}, respectively. The maximum outer radius of accretion flow is set at 5000 $R_\mathrm{S}$. For the ADAF solution, both the luminosity and radiation efficiency increase with the accretion rate until $\dot{m}_\mathrm{c}$ at the ISCO is reached. The luminosity of ADAF qualitatively has the trend $L_{\rm ADAF} \propto \dot{m}^2$, consistent with \citet{1995ApJ...452..710N}. Such a dependence of luminosity on the accretion rate can be understood as the variation of electron heating rate. With the increase in the accretion rate, the density increases and temperatures remain nearly constant. Thus, the heating rate to electrons increases as $q_{\rm ie}\propto n_{\rm e}^2 \propto \dot m^2$ (see Eq.(\ref{eq:qie})). The heating energy gained by electrons is all radiated in an optically thin ADAF, leading to the luminosity in the same dependence on the accretion rate, $L_{\rm ADAF} \propto \dot{m}^2$. An important consequence is that the radiation efficiency decreases with the decreasing accretion rate in a typical ADAF, unlike in the SSD it is a constant. Note that the direct viscous heating to electrons is not included in this work, which can exceed the collisional heating at very low accretion rates and hence results in a different dependence, $L_{\rm ADAF} \propto \dot m$.

Our generalized solution for the SSD has the same radiation efficiency $\eta_\mathrm{eff}=0.25$ as the theoretical value of SSD without the stress-free inner boundary assumption, which can be obtained from \citet{2002apa..book.....F}. With the increase of the accretion rate, the advection fraction of energy increases as photon trapping occurs, causing the radiation efficiency to decline from $\dot M\sim 0.1 \dot{M}_\mathrm{Edd}$ on. Then, the slim disc takes over the region within the photon trapping radius and expands outward with the further increase of the accretion rate. When the photon trapping radius reaches the outer boundary, the luminosity of the whole disc stops increasing with $\dot m$ since only the photons produced in surface layers can escape. These processes are generally consistent with previous investigations, e.g., \citet{2000PASJ...52..499M,2000PASJ...52..133W,2006ApJ...648..523W}.

\subsection{The spectrum}\label{subsec:spec}

\begin{figure}
	\includegraphics[width=\columnwidth]{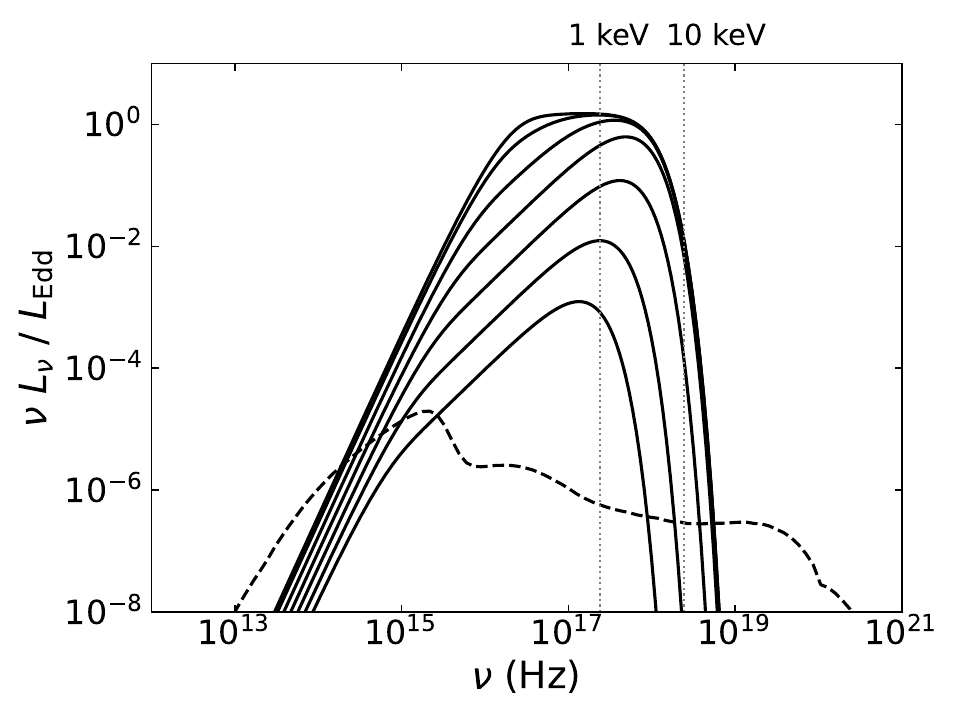}
    \caption{The spectra of the ADAF, SSD, and slim disc for $m=10$, $\alpha=0.1$, and $p_\mathrm{m}=p_\mathrm{g}$. The outer radius is set at 5000 $R_\mathrm{S}$ in the calculations. The dashed line is for ADAF with $\dot{m}$ = 0.001, the solid lines are for SSD-slim disc with $\dot{m}$ = 0.001, 0.01, 0.1, 1, 10, 100, and 1000 from bottom up.}
    \label{fig:spec}
\end{figure}

The spectra of SSD and slim disc are calculated by the multi-color blackbody with their effective temperatures, whereas the spectrum of ADAF is calculated through the Monte Carlo method \citep[see][]{1977AZh....54.1246P,1983ASPRv...2..189P,1997ApJ...489..791M,2012ApJ...744..145Q,2013arXiv1307.3635G}. The results are shown in Figure \ref{fig:spec}. The spectrum of ADAF for $\dot m=0.001$ is dominated by the synchrotron radiation at UV band, with bremsstrahlung and self-Compton scattering radiations in the X-ray band. The spectra of SSD are typical multi-color blackbody, extending to higher energy band and higher luminosity with the increase of the accretion rate. When the accretion rate exceeds a critical value, photon-trapping occurs at inner regions, the smaller distance, the more trapping. Thus, the peak of blackbody from different distances are accordingly cut down, forming a platform in the spectrum. The outer SSD region contributes multi-color blackbody to lower frequencies. The spectra of SSD-slim disc are consistent with \citet{2000PASJ...52..133W, 2019MNRAS.489..524K}. 

\section{Discussion} \label{sec:discussion}

In this work, we present a generalized solution describing various types of accretion flows. For simplicity, we adopt Newtonian gravity and only consider the Coulomb collisions for the heating of the electrons.

\begin{figure}
	\includegraphics[width=\columnwidth]{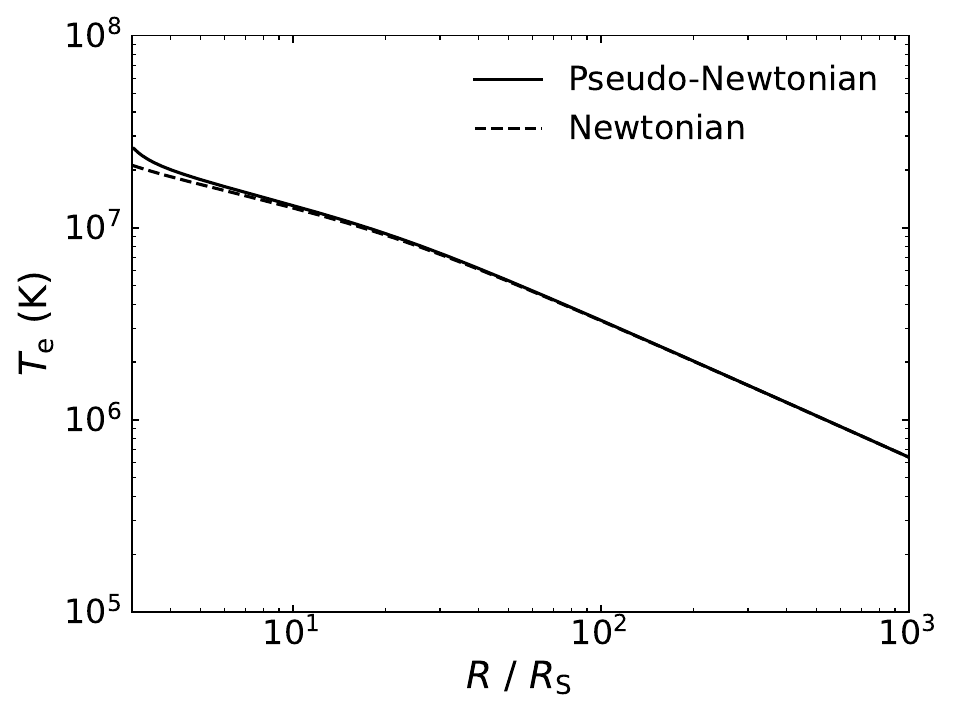}
    \caption{The radial profiles of electron temperature in Newtonian case (dashed line) and pseudo-Newtonian case (solid line) of the generalized solution for 10 $M_\odot$ Schwarzschild black hole, $\alpha=0.1$, and $p_\mathrm{m}=p_\mathrm{g}$, at 0.1 $\dot{M}_\mathrm{Edd}$.}
    \label{fig:GR}
\end{figure}

To evaluate the influence of relativistic effects, we perform the calculation under pseudo-Newtonian potential. The Keplerian rotational velocity in the generalized solution Eq.(\ref{eq:omega}) is modified as the \citet{1996ApJ...461..565A} form, for the dimensionless spin parameter $a_*\geq0$,
\begin{equation}
    \upsilon_\mathrm{K}=\Omega_\mathrm{K}R=\sqrt{\frac{GM}{R^{1-\beta_\mathrm{G}}\left(R-R_\mathrm{H}\right)^{\beta_\mathrm{G}}}},
\end{equation}
where
\vspace{-\baselineskip}

\begin{equation*}
    \begin{aligned}
        &R_\mathrm{H}=\frac{R_\mathrm{S}}{2}\left(1+\sqrt{1-a_*^2}\right),\\
        &R_\mathrm{ISCO}=\frac{R_\mathrm{S}}{2}\left[3+Z_2-\sqrt{\left(3-Z_1\right)\left(3+Z_1+2Z_2\right)}\right],\\
        &\beta_\mathrm{G}=\frac{R_\mathrm{ISCO}}{R_\mathrm{H}}-1,\\
        &Z_1\equiv1+\left(1-a_*^2\right)^{1/3}\left[\left(1+a_*^2\right)^{1/3}+\left(1-a_*^2\right)^{1/3}\right],\\
        &Z_2\equiv\sqrt{3a_*^2+Z_1^2}.
    \end{aligned}
\end{equation*}
As shown in Figure \ref{fig:GR}, the electron temperature only slightly rises near the ISCO with the pseudo-Newtonian modification for the Schwarzschild case. This indicates that the relativistic effects do not have a significant influence on this model for a Schwarzschild black hole. For the Kerr case, the general relativistic effects are generally significant, the luminosity and temperature in the inner zone are higher than those of the Newtonian case due to the small ISCO.

For the energy balance of electrons, we only include the Coulomb collisional heating and radiative cooling, neglecting the advected energy, compressive work, and direct viscous heating of the electrons. A complete energy equation of electrons is, $q_\mathrm{int,e}-q_\mathrm{c,e}=\delta q_\mathrm{vis}+q_\mathrm{ie}-q_\mathrm{rad}$, where $\delta$ is usually taken as $m_{\rm e}/m_{\rm i}\approx 10^{-3}$. The advected energy, compression work, and direct heat rate for electrons are compared with the radiative cooling as follows \citep[see e.g.,][]{1998ApJ...492..554N},
\begin{equation}
    \begin{aligned}
        & \frac{q_\mathrm{int,e}}{q_\mathrm{rad}}\approx\frac{2}{3\left(\Gamma_3-1\right)\epsilon_0\Xi}\cdot\frac{p_\mathrm{e}}{\beta p}\cdot\frac{f}{1-f}\sim \frac{p_\mathrm{e}}{\beta p}\cdot\frac{f}{1-f},\\
        & \frac{q_\mathrm{c,e}}{q_\mathrm{rad}}=\frac{1}{\epsilon_0\Xi}\cdot\frac{p_\mathrm{e}}{\beta p}\cdot\frac{f}{1-f}\sim \frac{p_\mathrm{e}}{\beta p}\cdot\frac{f}{1-f},\\
        & \frac{\delta q_\mathrm{vis}}{q_\mathrm{rad}}=\frac{1}{\Xi}\cdot\frac{\delta}{1-f}\sim \frac{\delta}{1-f}.
    \end{aligned}
\end{equation}
If all these ratios are much smaller than unity, the radiative cooling can only be balanced by the Coulomb collisional heating. For the SSD and the SLE, the fraction of advected energy $f$ is extremely small and the direct heating is negligible, thus, $q_\mathrm{ie}=q_\mathrm{rad}$ is a good approximation. For the slim disc, radiation pressure is much larger than the gas pressure of electrons, leading to the $q_\mathrm{int,e}$ and $q_\mathrm{c,e}$ unimportant. Although the viscosity heating would be important at high accretion rates, it cannot have any significant effects on the structure of the slim disc since only the Coulomb collision can produce a single-temperature accretion disc for super-Eddington accretion. For the ADAF, the advection and compressive heating can be the same order of magnitude as the radiative cooling only if $1-f$ is as small as $p_{\rm e}/p$, which corresponds to a very small accretion rate, $\dot m\la 0.1 \alpha^2\sim 10^{-3}$ \citep[see][]{1998tbha.conf..148N}. Therefore, $q_\mathrm{ie}=q_\mathrm{rad}$ is still a good approximation for $\dot m \ga 10^{-3}$. 

The general solution reveals that the innermost region of the accretion flow can become effectively optically thin, $\tau_\mathrm{eff}=(\tau\tau_\mathrm{abs})^{1/2}<1$, in the regime of moderate accretion rates between the SSD and slim disc. This usually occurs for supermassive black holes with large viscosity parameters at an accretion rate of $0.1-10$ Eddington rate. Such a solution has been mentioned by \citet{1973A&A....24..337S} and \citet{1973blho.conf..343N} but was only analyzed by a few researchers \citep[e.g.,][]{1998MNRAS.297..739B,2001ApJ...549.1050A,2004ApJ...614..101C, 2006ApJ...637..968A}. Our generalized solution provides a more precise description of this effectively optically thin solution. We will discuss the structure, spectrum, and stability of this solution in our following work (Liu et al. 2025 in prep.).

\section{Conclusion} \label{sec:conclusion}

We present an algebraic approach for the generalized equations describing the accretion flows for a wide range of accretion rates from sub- to super-Eddington. In the $f-\dot m$ plane, the ADAF, the SLE, the SSD, and the slim disc are well reproduced and unified by the generalized solution. The S-curve in $\dot{m}-\Sigma$ plane corresponding to SSD and slim disc branches is also well consistent with the early work of \citet{2002apa..book.....F} and \citet{1999ApJ...516..420W}. We reveal a hybrid radial structure of the optically thick accretion flows exist for a certain range of accretion rates, where the outer flow is consistent with the SSD \citep{2002apa..book.....F} and the inner flow is consistent with the slim disc \citep{1999ApJ...516..420W}, with a transition at a distance depending on the accretion rate. We calculate the radiations from the generalized solutions and show that the radiation efficiency increases with the accretion rate for the ADAF, stays constant for the SSD, and decreases again with photons trapped. The distinct spectra from the ADAF, the pure SSD, the radially SSD-slim disc, and the pure slim disc are presented according to the form of accretion flows at different accretion rates. In addition, we point out the existence of effectively optically thin accretion flow.

\section*{Acknowledgements}

We thank the anonymous referee for his/her useful comments. We acknowledge the support by the National Natural Science Foundation of China (Grant No.12333004). We also acknowledge the Beijing Super Cloud Center (BSCC) for providing HPC resources.

\section*{Data Availability}

The code and data underlying this article will be shared on reasonable request to the corresponding author.







\appendix

\section{The heating and cooling of electrons} \label{app:model}
The energy transfer rate between ions and electrons, $q^{\mathrm{ie}}$, is given by \citet{1983MNRAS.204.1269S},
\begin{equation}\label{eq:qie}
    \begin{aligned}
        q^{\mathrm{ie}}&=\frac{3m_\mathrm{e}}{2m_\mathrm{p}}\sum_j Z^2_jn_jn_\mathrm{e} \sigma_\mathrm{T}c\frac{k_\mathrm{B}(T_\mathrm{i}-T_\mathrm{e})}{K_2(1/\theta_\mathrm{i})K_2(1/\theta_\mathrm{e})}\ln{\Lambda}\\
        &\times\left[\frac{2(\theta_\mathrm{i}+\theta_\mathrm{e})^2+1}{\theta_\mathrm{i}+\theta_\mathrm{e}}K_1\left(\frac{\theta_\mathrm{i}+\theta_\mathrm{e}}{\theta_\mathrm{i}\theta_\mathrm{e}}\right)+2K_0\left(\frac{\theta_\mathrm{i}+\theta_\mathrm{e}}{\theta_\mathrm{i}\theta_\mathrm{e}}\right)\right],
    \end{aligned}
\end{equation}
where $m_\mathrm{p}$ and $m_\mathrm{e}$ are the mass of proton and electron, respectively, $Z_j$ is the charge number of the $j$-th species, $n_j$ is the number density of the $j$-th species, i.e., $\sum_j n_j=n_\mathrm{i}$, the Coulomb logarithm $\ln\Lambda$ is taken as 20 in this work, $\theta_\mathrm{i}=k_\mathrm{B}T_\mathrm{i}/m_\mathrm{i}c^2$ and $\theta_\mathrm{e}=k_\mathrm{B}T_\mathrm{e}/m_\mathrm{e}c^2$ are the dimensionless ion and electron temperatures, respectively, $K_n$ is the $n$-th order modified Bessel function.

The bremsstrahlung radiation is mainly produced by the electron-ion and electron-electron interactions when the accretion rate is not extremely high so that the pair production process can be ignored. We adopt the expression of bremsstrahlung cooling rate per unit volume following \citet{1982ApJ...258..335S,1983MNRAS.204.1269S,1995ApJ...452..710N},
\begin{equation}
    \begin{split}
        q_\mathrm{br} &=q_\mathrm{ei}+q_\mathrm{ee}, \\
        q_\mathrm{ei} &=\sum_j Z^2_jn_jn_\mathrm{e}\sigma_\mathrm{T}c\alpha_f m_\mathrm{e}c^2F_\mathrm{ei}\left(\theta_\mathrm{e}\right),\\
        q_\mathrm{ee} &=n_\mathrm{e}^2r_\mathrm{e}^2c\alpha_f m_\mathrm{e}c^2F_\mathrm{ee}\left(\theta_\mathrm{e}\right),
    \end{split}
\end{equation}
where
\vspace{-\baselineskip}

\begin{align*}
    F_\mathrm{ei}\left(\theta_\mathrm{e}\right)&=
    \begin{cases}
    \displaystyle{4\left(\frac{2\theta_\mathrm{e}}{\pi^3}\right)^{1/2}\left(1.781+\theta_\mathrm{e}^{1.34}\right)}, & \theta_\mathrm{e}<1,\\
    \displaystyle{\frac{9\theta_\mathrm{e}}{2\pi}\left[\ln{\left(1.123\theta_\mathrm{e}+0.48\right)+1.5}\right]}, & \theta_\mathrm{e}>1,
    \end{cases}\\
    F_\mathrm{ee}\left(\theta_\mathrm{e}\right)&=
    \begin{cases}
    \begin{aligned}
        &\displaystyle{\frac{20}{9\pi^{1/2}}\left(44-3\pi^2\right)\theta_\mathrm{e}^{3/2}}\\
        &\times\left(1+1.1\theta_\mathrm{e}+\theta_\mathrm{e}^2-1.25\theta_\mathrm{e}^{5/2}\right)
    \end{aligned}, & \theta_\mathrm{e}<1,\\
    24\theta_\mathrm{e}\left(\ln{2\eta_\mathrm{E}\theta_\mathrm{e}}+1.28\right), & \theta_\mathrm{e}>1
    \end{cases}
\end{align*}
with $\alpha_f$ the fine-structure constant, $r_\mathrm{e}=e^2/m_\mathrm{e}c^2$ the classical radius of electrons, $e$ the electron charge, $\eta_\mathrm{E}=\exp{\left(-\gamma_\mathrm{E}\right)}\approx0.5616$, and $\gamma_\mathrm{E}$ the Euler constant.

The synchrotron cooling rate is estimated by the approximation \citep{1995ApJ...452..710N},
\begin{equation}
    q_\mathrm{syn}\approx\frac{2\pi}{3c^2}k_\mathrm{B}T_\mathrm{e}\frac{\nu_\mathrm{syn}^3}{R},
\end{equation}
where $\nu_\mathrm{syn}$ is the critical frequency, $\nu_\mathrm{syn}=3\nu_0\theta_\mathrm{e}^2x_M/2$, $\nu_0\equiv eB/2\pi m_\mathrm{e}c$, the coefficient $x_M$ is estimated as $\lg x_M=3.6+0.25\lg\dot{m}$ following \citet{1997ApJ...477..585M}, and the Eddington scaled accretion rate $\dot{m}$ is defined in Section \ref{sec:results}.

The Compton cooling rates $q_\mathrm{br,c}$ and $q_\mathrm{syn,c}$ are calculated through the average luminosity enhancement factor $\eta$, following \citet{1991ApJ...369..410D} and \citet{1995ApJ...452..710N},
\begin{equation}
    \eta=1+\frac{P(A-1)}{1-PA}\left[1-\left(\frac{x}{3\theta_\mathrm{e}}\right)^{-1-\ln P/\ln A}\right],
    \label{eq:eta}
\end{equation}
where $P=1-\exp(-\tau_\mathrm{es})$ is the scattering probability of photons and $A=1+4\theta_\mathrm{e}+16\theta_\mathrm{e}^2$ is the mean amplification factor of photons in single scattering. Eq.(\ref{eq:eta}) treats both the unsaturated and saturated limits of Comptonization and provides a generalization of the result derived by \citet{Kompaneets1957} for the Comptonization of bremsstrahlung photons \citep{1991ApJ...369..410D}.
 
Therefore, the flat spectrum of bremsstrahlung from the self-absorption edge to the high-energy cutoff ($h\nu\approx k_{\rm B}T_\mathrm{e}$), $q_{\rm br}(\nu)=(q_{\rm br}h/k_{\rm B}T_{\rm e})\exp{\left(-h\nu/k_{\rm B}T_{\rm e}\right)}\approx q_{\rm br}{h/k_{\rm B}T_{\rm e}}$, is amplified by a factor of $\eta$ and the net cooling rate can be evaluated as
\begin{equation}
q_\mathrm{br,C}={\frac{q_\mathrm{br}}{{\theta_\mathrm{e}}}}\int_{x_\mathrm{c}}^{\theta_\mathrm{e}}\left(\eta-1\right) \text dx,
\end{equation}
where $x=h\nu/m_\mathrm{e}c^2$ is the dimensionless frequency of photons and $x_\mathrm{c}\equiv h\nu_\mathrm{c}/m_\mathrm{e}c^2$. The self-absorption frequency $\nu_\mathrm{c}$ is taken as the critical frequency of synchrotron $\nu_\mathrm{syn}$. 
The Comptonization of synchrotron radiation is concentrated on the critical frequency $\nu_\mathrm{syn}$,
\begin{equation}
 q_\mathrm{syn,C}=\left(\eta-1\right)q_\mathrm{syn}.
\end{equation}

\section{Why dose the self-similar approximation work?}
\label{app:solution}

To understand the self-similar approximation, we transform the general equations and show that all the radial derivatives can be re-written in their logarithmic forms. The radial momentum equation (\ref{eq:radial}) is re-written as,
\begin{equation}\label{eq:int_radial}
   -\upsilon^2\frac{\text d\ln{\upsilon}}{\text d\ln{R}}+\left(\Omega R\right)^2-c_\mathrm{s}^2\frac{\text d\ln{p}}{\text d\ln{R}}=\upsilon^2_{\text K}.
\end{equation}

Combined with the continuity equation (\ref{eq:continuity}) and viscosity law $\nu=\alpha c_\mathrm{s}^2/\Omega_\mathrm{K}$, the azimuthal momentum equation (\ref{eq:angular}) can be integrated from ISCO to a distance $R$, which gives the relation of the velocities as, 
\begin{equation}\label{eq:int_angular2}
    \upsilon=\frac{\alpha c_\mathrm{s}^2}{\Omega_\mathrm{K}R}\frac{\text d\ln{\Omega}}{\text d\ln{R}}g^{-1},
\end{equation}
where $g=1-\Omega (R_{\rm in})R_{\rm in}^2/\Omega(R)R^2$, $g=1$ if the inner boundary neglected.

The energy equation (\ref{eq:energy}), by adopting a fraction of energy advected and using the integration of Eq.(\ref{eq:angular}), can be transformed to
\begin{equation}\label{eq:int_energy}
    \frac{c_\mathrm{s}^2}{\Gamma_3-1}\frac{\text d}{\text d\ln{R}}\left(\ln{\frac{c_\mathrm{s}^2}{\Gamma_3-1} }\right)-fc_\mathrm{s}^2\frac{\text d\ln{\rho}}{\text d\ln{R}}=f\left(\Omega R\right)^2\frac{\text d\ln{\Omega}}{\text d\ln{R}}g.
\end{equation}

Eq.(\ref{eq:int_radial}), (\ref{eq:int_angular2}), and (\ref{eq:int_energy}) combined with Eq.(\ref{eq:continuity}) and (\ref{eq:vertical}) are the new forms of the general equations including variables $\upsilon(R)$, $\Omega(R)$, $c_s(R)$, $\rho(R)$, and $H(R)$. As long as these variables have a power-law form with respect to $R$, the derivatives turn out to be the power indices. In the self-similar solutions, specific power indices are chosen on the basis of intrinsic physics, deviation from the true indices in general solutions (including SSD and slim disc) is not so large to significantly change the solutions, in particular, taking into account the approximation included in the vertical hydrostatic equilibrium and the uncertainty in viscous parameter and magnetic parameter. Moreover, we perform the iterative calculations to determine the coefficients of the power-law variables (such as $c_1$ and $c_2$ of the velocities in Eq.\ref{eq:omega}) for every grid point of the distance, which modified the power indices of the self-similar solutions in some extent. The computational results further confirm the self-similar approximation.

\section{The thermal stability of the generalized solution}
\label{app:stable}

We introduce a new method to analyse the locally thermal stability of our generalized solution, as well as those of the four basic solutions, emphasizing the effect of advection. To achieve thermal stability in the accretion flow, it is required that 
\begin{equation} \label{eq:thermal stable criter}
    \frac{\text d(1-f)Q^+}{\text dT}<\frac{\text dQ^-}{\text dT},
\end{equation}
where $Q^+=(q_\mathrm{vis}+q_\mathrm{c})2H$ is the total energy dissipation per unit surface area and $Q^-=2F_\mathrm{rad}$ is the column radiative energy.

We begin by estimating the structural parameters via the conservation equations given in Section \ref{subsec:equation}. The energy equation (\ref{eq:energy}) can be rewritten as $q_\mathrm{int}=f(q_\mathrm{vis}+q_\mathrm{c})=f\Xi q_\mathrm{vis}$, where $q_\mathrm{int}=-\rho\upsilon c_\mathrm{s}^2/(\Gamma_3-1)R$ is the advected energy and $q_\mathrm{vis}=9\nu\rho\Omega^2/4$ is the viscous heating rate. The radial velocity can be expressed as $\upsilon=\nu\text d\ln{\Omega}/\text dR\approx-3\nu/2R$ according to Eq.(\ref{eq:int_angular2}). The angular velocity is given by $\Omega=c_2\Omega_\mathrm{K}$, where the coefficient $c_2$ is of order unity \citep[see e.g. Eq.\ref{eq:omega} and][]{1994ApJ...428L..13N,1995ApJ...452..710N}. Eliminating $\upsilon$, $\nu$, and $\Omega$ from the energy equation, we can obtain $c_\mathrm{s}^2\approx3c^\prime f\Xi(\Omega_\mathrm{K}R)^2/2$, where the coefficient $c^\prime=c_2^2\left(\Gamma_3-1\right)$ is also of order unity for most cases. Then, $\upsilon$, $\Omega$, $c_\mathrm{s}$, $H$, $\Sigma$, $\rho$, $p$, $\tau_\mathrm{es}$, $q_\mathrm{vis}$, and $Q^+$ can be estimated as follows with the equations in Section \ref{subsec:equation}, 
\begin{equation}\label{eq:estimation}
    \begin{split}
        &\upsilon\approx-4.77\times10^{10}c^\prime f\Xi\alpha r^{-1/2}\ \text{cm}\ \text{s}^{-1},\\
        &\Omega=7.18\times10^4c_2 m^{-1}r^{-3/2} \ \text{s}^{-1},\\
        &c_\mathrm{s}^2\approx6.74\times10^{20}c^\prime f\Xi r^{-1}\ \text{cm}^2\ \text{s}^{-2},\\
        &H\approx5.72\times10^5\left(c^\prime f\Xi\right)^{1/2}mr\ \text{cm},\\
        &\Sigma\approx15.7\left(c^\prime f\Xi\right)^{-1}\alpha^{-1}\dot{m}r^{-1/2}\ \text{g}\ \text{cm}^{-2},\\
        &\rho\approx1.37\times10^{-5}\left(c^\prime f\Xi\right)^{-3/2}\alpha^{-1}m^{-1}\dot{m}r^{-3/2}\ \text{g}\ \text{cm}^{-3},\\
        &p\approx9.26\times10^{15}\left(c^\prime f\Xi\right)^{-1/2}\alpha^{-1}m^{-1}\dot{m}r^{-5/2}\ \text{g}\ \text{cm}^{-1}\ \text{s}^{-2},\\
        &\tau_\mathrm{es}\approx2.75\left(c^\prime f\Xi\right)^{-1}\alpha^{-1}\dot{m}r^{-1/2},\\
        &q_\mathrm{vis}\approx1.50\times10^{21}c_2^2\left(c^\prime f\Xi\right)^{-1/2}m^{-2}\dot{m}r^{-4}\ \text{erg}\ \text{cm}^{-3}\ \text{s}^{-1},\\
        &Q^+\approx1.71\times10^{27}c_2^2\Xi m^{-1}\dot{m}r^{-3}\ \text{erg}\ \text{cm}^{-2}\ \text{s}^{-1}.
    \end{split}
\end{equation}

\begin{figure*}
	\includegraphics[width=\textwidth]{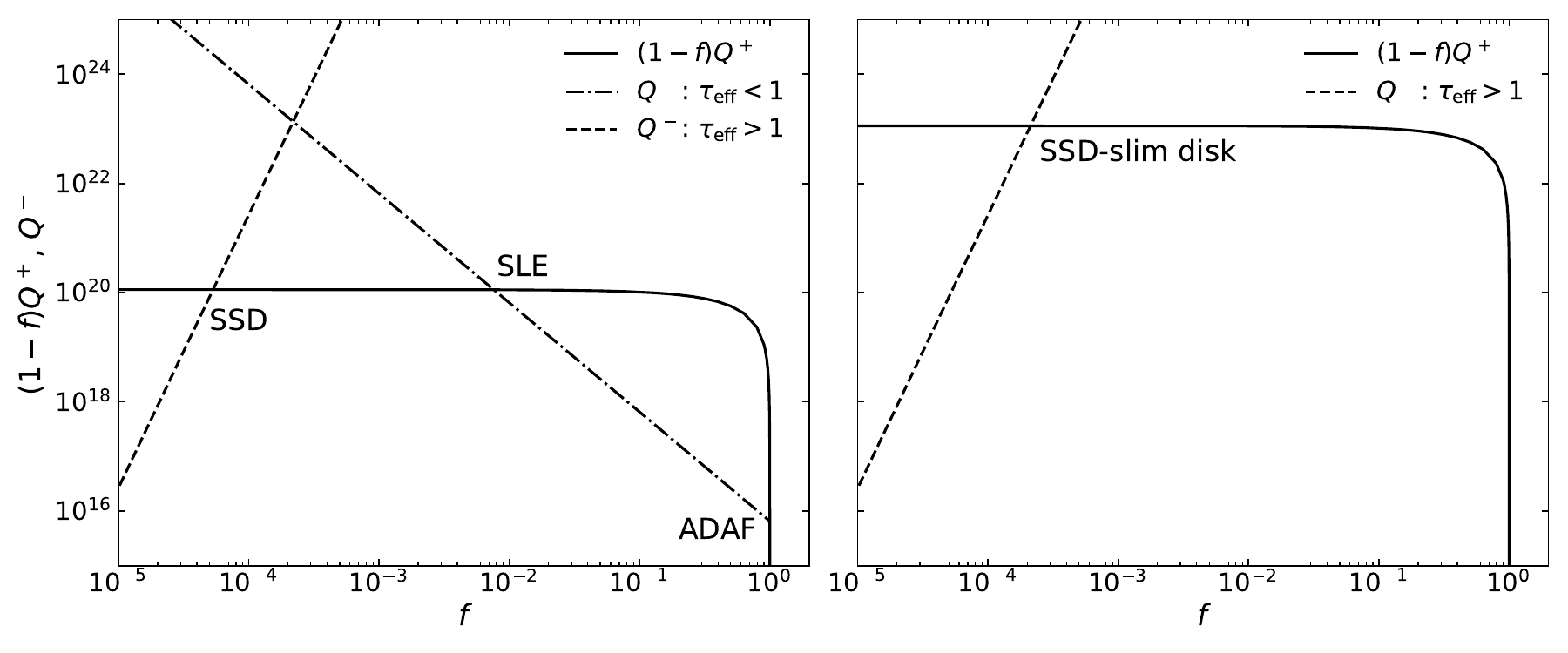}
    \caption{The schematic diagram of heating and cooling rates as functions of advected fraction $f$ for gas pressure supported accretion flow (left panel) and radiation pressure supported accretion flow (right panel). We take $\alpha=0.1$, $\beta=0.5$, $m=10$, $r=10$, $\dot{m}=10^{-3}$ (left panel) and $\dot{m}=1$ (right panel) as an example. Each intersection point of $(1-f)Q^+$ and $Q^-$ represents a state of thermal equilibrium of the system and hence corresponds to one solution of the accretion flow. The intersection point of the two $Q^-$ curves in the left panel corresponds to $\tau_\mathrm{eff}=1$. In the calculation of this schematic diagram we have adopted $\Lambda=1$, $\Xi=1$, $\varepsilon=1$, $\xi=0.01$ for simplicity, but these curves are still qualitatively consistent with the heating and cooling rates in our numerical results and \citet{1996ApJ...465..312E}.}
    \label{fig:stable}
\end{figure*}

To obtain $Q^-$, we need to estimate the temperature $T$ from EOS (\ref{eq:p}) for different accretion flows. In the gas-pressure-supported SSD, the SLE, and the ADAF, we roughly express the gas pressure as $\beta\rho c_\mathrm{s}^2\approx (n_\mathrm{i}+n_\mathrm{e})k_\mathrm{B}T$. In this estimation, we effectively neglected the contribution of electrons to the gas pressure, which is accurate to at least a factor of two since $T_\mathrm{i}\geq T_\mathrm{e}$. For the radiation-pressure-supported SSD and slim disc, the accretion flow is almost one temperature, giving $p=4\sigma T^4/3c$. Therefore, temperature and thermal stable criterion Eq.(\ref{eq:thermal stable criter}) can be expressed as, 
\begin{equation}\label{eq:criter}
    \begin{aligned}
        &T\approx4.80\times10^{12}c^\prime f\Xi\beta r^{-1}\ \text{K},\\
        &T\approx4.38\times10^7\left(c^\prime f\Xi\right)^{-1/8}\alpha^{-1/4}m^{-1/4}\dot{m}^{1/4}r^{-5/8}\ \text{K},
    \end{aligned}
\end{equation}
for the gas-pressure-dominated accretion flow and optically thick radiation-pressure-dominated accretion flow, respectively. The former corresponds to $\text d(1-f)Q^+/\text df<\text dQ^-/\text df$, while the latter is the opposite.

Now we can obtain $Q^-$ by analysing the vertical radiative transfer. In optically thick flows (SSD and slim disc), the radiative diffusion approximation gives $Q^-=16\sigma T^4/3\tau$. For the optically thin accretion flow, the emergent radiation flux can be simply estimated to be the total emissivity $F_\mathrm{rad}\approx(q_\mathrm{br}+q_\mathrm{syn}+q_\mathrm{br,C}+q_\mathrm{syn,C})H$. Besides, $Q^-$ of the radiation pressure supported accretion flow can be directly estimated as $Q^-\approx4cp/\tau$, through the expression of radiation pressure in Eq.(\ref{eq:p}). For convenience, we introduce two new notations, $\xi\equiv T_\mathrm{e}/T_\mathrm{i}$ and $\Lambda\equiv(q_\mathrm{br}+q_\mathrm{syn}+q_\mathrm{br,C}+q_\mathrm{syn,C})/q_\mathrm{br}$, where $q_\mathrm{br}=5.43\times10^{20}\rho^2T^{1/2}\xi^{1/2}\ \text{erg}\ \text{cm}^{-3}\ \text{s}^{-1}$ assuming the Gaunt factor is $\bar{g}_\mathrm{B}=1.2$ \citep{Rybicki_Lightman_1979}. We also adopt the probability for scattering $1-\varepsilon\equiv\tau_\mathrm{es}/\tau$ from \citet{Rybicki_Lightman_1979}. The column radiative energy $Q^-$ can then be expressed as,
\begin{equation}
    \begin{aligned}
        &Q^-\approx2.57\times10^{23}\xi^{1/2}\Lambda\left(c^\prime f\Xi\right)^{-2}\alpha^{-2}\beta^{1/2}m^{-1}\dot{m}^2r^{-5/2},\\
        &Q^-\approx5.85\times10^{46}\left(1-\varepsilon\right)\left(c^\prime f\Xi\right)^5\alpha\beta^4\dot{m}^{-1}r^{-7/2},\\
        &Q^-\approx4.03\times10^{26}\left(1-\varepsilon\right)\left(c^\prime f\Xi\right)^{1/2}m^{-1}r^{-2},
    \end{aligned}
\end{equation}
in units of erg cm$^{-2}$ s$^{-1}$, for optically thin gas-pressure dominated accretion flow, optically thick gas-pressure dominated accretion flow, and radiation-pressure-dominated accretion flow, respectively.

Now that we have found the expressions of $Q^-$ and $Q^+$ in terms of the advection fraction $f$, the energy balance $(1-f)Q^+=Q^-$ in all four basic solutions of the accretion flow can be written as equations of $f$,
\begin{equation}\label{eq:stable}
    \begin{aligned}
    1-f&=1.50\times10^{-4}c_2^{-2}c^{\prime-2}\xi^{1/2}\Lambda f^{-2}\Xi^{-3}\alpha^{-2}\beta^{1/2}\dot{m}r^{1/2},\\ 
    1-f&=3.42\times10^{19}\left(1-\varepsilon\right)c_2^{-2}c^{\prime5}f^{5}\Xi^{4}\alpha\beta^4 m\dot{m}^{-2}r^{-1/2},\\ 
    1-f&=0.24\left(1-\varepsilon\right)c_2^{-2}c^{\prime1/2}f^{1/2}\Xi^{-1/2}\dot{m}^{-1}r,
    \end{aligned}
\end{equation}
for optically thin gas-pressure-dominated accretion flow, optically thick gas-pressure-dominated accretion flow, and radiation-pressure-dominated accretion flow, respectively.

When evaluated in a $Q-f$ plane as shown in Figure \ref{fig:stable} (left panel), it is clear that the first equation of Eq.(\ref{eq:stable}) (gas pressure supported optically thin flow), has two solutions at most, corresponding to the ADAF and SLE solutions respectively. One can also notice from the $Q-f$ plane that the ADAF solution satisfies the thermal stable criterion whereas the SLE solution does not. Furthermore, since the radiative cooling is more sensitive to the accretion rate than the viscous dissipation ($q_\mathrm{br}\propto\dot{m}^2$, $q_\mathrm{vis}\propto\dot{m}$), the radiative cooling would become so efficient that the gas can no longer remain a hot accretion flow when the accretion rate rises above a certain critical value, $\dot{m}_\mathrm{c}$, which is exactly the critical accretion rate of ADAF discussed repeatedly in literature. In the above analysis, $\dot{m}_\mathrm{c}$ corresponds to the value when the first equation has only one solution, i.e., when the ADAF solution and SLE solution merge. It can be derived that for this condition to be satisfied, $f=2/3$ is required. We can thus obtain the qualitative expression of the critical accretion rate as,
\begin{equation}
    \dot{m}_\mathrm{c}\approx0.01\frac{C}{10^{-3}}\left(\frac{T_\mathrm{i}}{T_\mathrm{e}}\right)^{1/2}\Lambda^{-1}\Xi^3\left(\frac{\alpha}{0.1}\right)^2\beta^{-1/2}r^{-1/2},
    \label{eq:mdotc}
\end{equation}
where the coefficient $C=c_2^2c^{\prime2}=c_2^6\left(\Gamma_3-1\right)^2$ has a magnitude of a few $10^{-3}$, $c_2^2=(4-3\beta)/(9-3\beta)$, and $\Gamma_3=(8-3\beta)/(6-3\beta)$ for $p_\mathrm{g}>p_\mathrm{r}$.

Again from Figure \ref{fig:stable} (left panel), it can be straightforwardly demonstrated that the second equation of Eq.(\ref{eq:stable}) (gas pressure supported optically thick flow) has only one physical solution. Moreover, since $(1-f)Q^+$ has a negative correlation with $f$, whereas the cooling rate has a positive one, the thermal stability of gas-pressure-supported SSD is naturally explained. In addition, when the accretion rate is lower than another critical value, $\dot{m}^\prime_\mathrm{c}$, the low density would lead to a radiative cooling so inefficient that the gas cannot remain cold and optically thick. Similar to the $\dot{m}_\mathrm{c}$ discussed above, $\dot{m}^\prime_\mathrm{c}$ corresponds to the value when the SLE solution and SSD solution merge, i.e. the three curves in the left panel of Figure \ref{fig:stable} intersect at one single point.

Similar to the analysis of the first two equations of Eq.(\ref{eq:stable}), it can be shown from the right panel of Figure \ref{fig:stable} that the third equation (the radiation pressure supported optically thick flow) also has only one solution, corresponding to the radiation-pressure-supported SSD and the slim disc. Expectedly, the thermal stable criterion, Eq.(\ref{eq:criter}), shows that this solution is locally thermally unstable, although for slim disc, this instability can be suppressed by advection, leading to a globally stable solution \citep[the detailed analyses can be seen in the discussions, e.g.,][]{1991PASJ...43..147H,1991PASJ...43..261H,1997MNRAS.287..165S,1998MNRAS.298..888S,2001MNRAS.328...36S,2008bhad.book.....K}. Moreover, the viscous timescale is comparable to the thermal timescale in slim disc, which means the steady-state continuity equation and azimuthal momentum equation are no longer valid for the analysis of the thermal stability of slim disc.

We leave a detailed discussion of the radiation pressure supported optically thin solution mentioned in Section \ref{sec:discussion} and the analysis of its thermal stability for our next work.


\bsp	
\label{lastpage}
\end{document}